\documentclass[acmlarge]{acmart}
\usepackage{url}
\usepackage{booktabs} 
\usepackage{listings}
\usepackage{algorithmic}
\usepackage{graphics}   
\usepackage{comment}
\usepackage{colortbl}
\usepackage[final]{changes}
\usepackage{subcaption}
\usepackage[ruled]{algorithm2e} 

\SetAlFnt{\small}
\SetAlCapFnt{\small}
\SetAlCapNameFnt{\small}
\SetAlCapHSkip{0pt}
\IncMargin{-\parindent}

\begin{abstract}
\acmJournal{JOCCH}
\acmVolume{9}
\acmNumber{4}
\acmArticle{39}
\acmYear{2010}
\acmMonth{3}
\acmArticleSeq{11}
\end{abstract}

\setcopyright{usgovmixed}

\acmDOI{0000001.0000001}

\received{February 2007}
\received{March 2009}
\received[accepted]{June 2009}
\graphicspath{ {graphics/} }

\begin{document}
\title{Understanding \replaced{Group Event Scheduling }{Group Event Dynamics} via the OutWithFriendz Mobile Application} 

\author{Shuo Zhang}
\orcid{1234-5678-9012-3456}
\affiliation{%
  \institution{University of Colorado Boulder}
  \streetaddress{1111 Engineering Dr}
  \city{Boulder}
  \state{CO}
  \postcode{80309}
  \country{USA}}
\author{Khaled Alanezi}
\affiliation{%
  \institution{University of Colorado Boulder}
  \department{1111 Engineering Dr}
  \city{Boulder}
  \state{CO}
  \postcode{80309}
  \country{USA}
}
\author{Mike Gartrell}
\affiliation{%
  \institution{University of Colorado Boulder}
  \department{1111 Engineering Dr}
  \city{Boulder}
  \state{CO}
  \postcode{80309}
  \country{USA}
}
\author{Richard Han}
\affiliation{%
  \institution{University of Colorado Boulder}
  \department{1111 Engineering Dr}
  \city{Boulder}
  \state{CO}
  \postcode{80309}
  \country{USA}
}
\author{Qin Lv}
\affiliation{%
  \institution{University of Colorado Boulder}
  \department{1111 Engineering Dr}
  \city{Boulder}
  \state{CO}
  \postcode{80309}
  \country{USA}
}
\author{Shivakant Mishra}
\affiliation{%
  \institution{University of Colorado Boulder}
  \department{1111 Engineering Dr}
  \city{Boulder}
  \state{CO}
  \postcode{80309}
  \country{USA}
}

\begin{abstract}
The wide adoption of smartphones and mobile applications has brought significant changes to not only how individuals behave in the real world, but also how groups of \deleted{mobile} users interact with each other when organizing group events. Understanding how \replaced{users make event decisions as a group }{mobile groups make event decisions} and identifying the contributing factors can offer important insights for social group studies and more effective system and application design for \replaced{group event scheduling }{groups}. 

In this work, we have designed a new mobile application called OutWithFriendz, which enables \replaced{users of our mobile app }{groups of mobile users} to organize group events, invite friends, suggest and vote on event time and venue. We have deployed OutWithFriendz at both Apple App Store and Google Play, and conducted a large-scale user study spanning over 500 users and 300 group events. Our analysis has revealed several important observations regarding group event \replaced{planning s}{decision} process including the importance of user mobility, individual preferences, host preferences, and group voting process. 

\end{abstract}

%
%


\maketitle

\renewcommand{\shortauthors}{S. Zhang et. al.}

\section{Introduction}
\label{sec:intro} 

The ability of users to organize events using mobile devices is a defining 
characteristic of today's social network systems. With the advancement of 
mobile technology, more and more people are digitally connected, which 
makes \added{the} analysis of \replaced{group event planning process and decision making }
{group behavior and suggestions}  for event organizers critically important. 
There is a rich history of UbiComp research concentrating on individual 
user behavior analysis, which treats individual user's data as a singleton.
However, social interactions among group members are often ignored, and
relatively scant research to date has explored the subject of 
\replaced{group event scheduling }{mobile group dynamics}. 
Some early work~\cite{beckmann2011agremo, park2008restaurant} has 
been confined to in-lab surveys and has not studied \added{the} real-world
\replaced{event scheduling process by}{dynamics of mobile}  
 groups of users, nor \replaced{the factors that would }{its} impact
\deleted{on} group \added{event} decision-making.  More recently, a study of university groups
using mobile phones was presented~\cite{jayarajah2015need}.  What has been missing to date is a detailed
understanding of the \replaced{process }{dynamics} of how groups 
make a decision to visit a particular place at a particular time \added{using their mobile devices}. 
What factors influence a group's final decision? This paper provides detailed novel insights in the \replaced{event scheduling }{decision-making} process of \replaced{social }{mobile} groups.  
We believe that this is an exciting area ripe for exploration by the ubiquitous computing research 
community. With ever-increasing popularity of smartphones, we 
expect that mobile computing will be used extensively to assist
groups of people in \replaced{event planning }{making decisions}
\replaced{in terms of}{about} when and where to rendezvous. Consider a group of
friends out on a weekend evening trying to decide what movie to see or where to eat, or consider a group of professional
colleagues trying to decide where to go for lunch. \added{Given the frequency with which people
schedule colocated events,} 
we believe mobile applications \added{for group event scheduling} can provide significant help.

\added{However, despite this considerable potential, today's technology offers limited help when it 
comes to coordinating group events in online and offline scenarios.} 
Currently there are few group event organization applications on the market. 
The most commonly used services are Meetup~\cite{meetup-about}, 
Facebook Events~\cite{facebook} and Evite~\cite{evite}. 
In these services, hosts organize offline events and post them on the website.
Users or group members who are interested in these events RSVP and later
attend the events in real life. However, in all of these existing services, meeting 
time and location are settled by the host at the creation of the event. Potential
users are not able to sufficiently express their opinions on when and where to
meet. This will more or less have a negative impact on event attendance. 
Doodle~\cite{doodle} is an online event scheduling service which supports 
groups in finding a mutually agreeable meeting time. Participants are able to vote 
for their \added{time} preferences. But all the meeting time options are \replaced{pre-selected by the host }
{added by host in the beginning}. Group members have no permission to suggest new options. In addition,
group members cannot suggest or vote for meeting locations. 

To address the limitations of existing services, we developed OutWithFriendz, a mobile
application that enables groups of people to decide together through a voting
process the date/time the group would like to meet as well as the location where 
they would like to meet. OutWithFriendz is implemented as a client-server 
architecture that is comprised of both iOS and Android based clients that 
communicate with a server implemented as a Java Web application.

\begin{figure}[t] 
\centering
\includegraphics[width=0.9\linewidth]{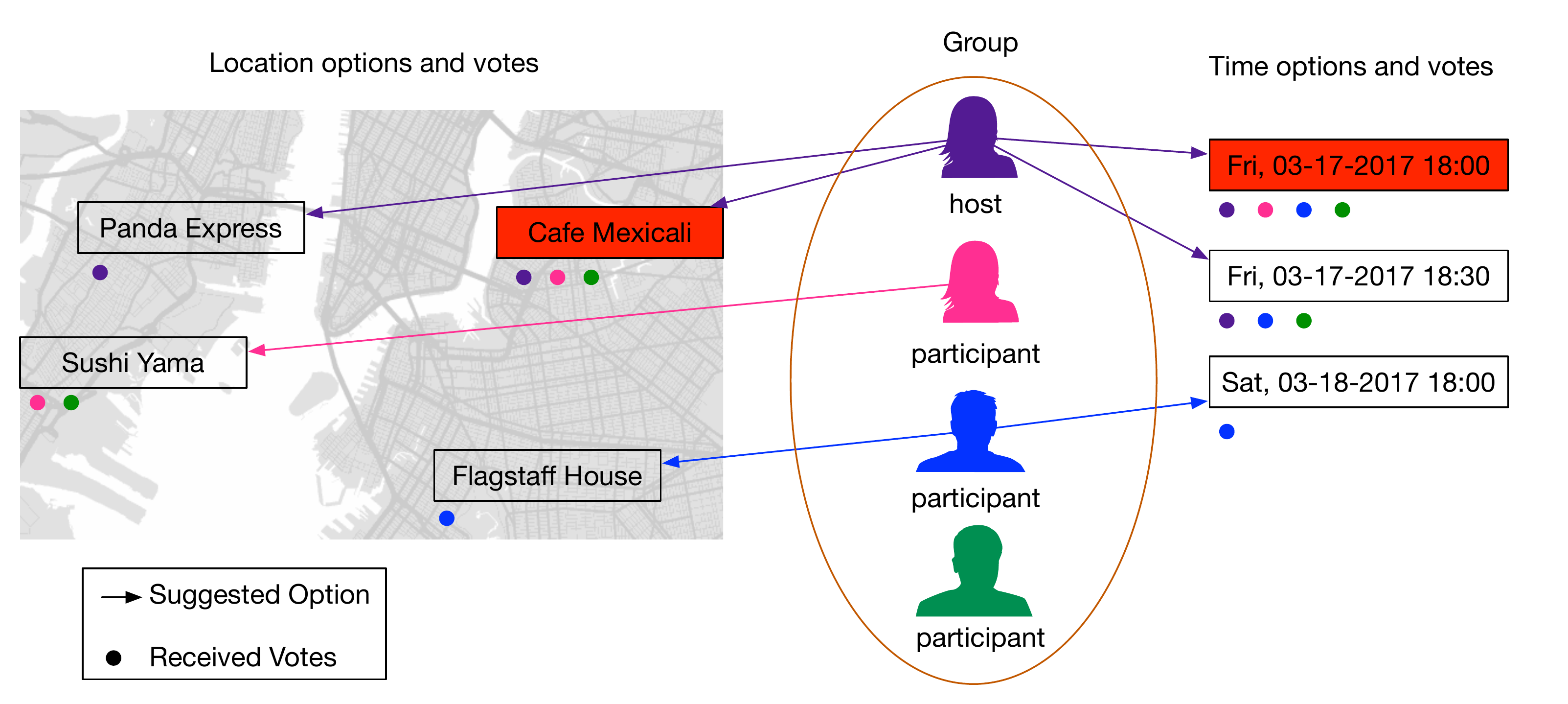}
\caption{An illustration of the key elements in our OutWithFriendz system. \added{The colored arrows (dots) indicate
which user suggested (voted) for a meeting time or location, and the red boxes indicate the final decisions.}}
\label{fig:intro}
\end{figure}

The main elements of our OutWithFriendz mobile application are shown in 
Figure~\ref{fig:intro}. To start using it, a user may create a new invitation acting
as a host. During this process, she can specify the details of this invitation including
a title, a list of suggested dates, a list of suggested locations and invited participants.
After this host submits a new invitation to the server, all invited participants
receive it and can view the detailed invitation information on their own clients. They 
can then suggest more dates, locations or vote for their preferred options and comment 
on the invitation. \added{After the voting process has ended, } the host then
decides on the final meeting location and time, whuch are then sent to all participants. 
In this example (Figure~\ref{fig:intro}), the host suggested four locations and three date/time options. After the suggesting and voting process, she selected Cafe Mexicali and Friday, 03-17-2017, 18:00 as final decisions, which received 
the most votes. Please note that, in our design, the host can make decisions based on the 
voting results, but she does not have to always obey them. We do find few hosts in our field study
whose final decisions was different from the options that received most votes.
This scenario will be discussed later.

Introducing our newly designed OutWithFriendz mobile application which embeds group
decision making into the voting process raises new questions: How do the mobile \added {app} users collaborate
to organize their group events? What are the major factors that will impact group decisions?
How is the voting behavior processed? And how to improve the event attendance rate?

Our mobile application also collects user mobility-related data. The app posts GPS user 
location traces to the server. Users may opt out from providing their location traces, although
most users did not disable location tracking for the entire duration of their participation in the
study. This user mobility data provides great opportunity to derive such input factors as spread,
movement, mobility, and to investigate their impact on group \replaced{event scheduling }{decisions}.

The contributions of this paper can be summarized as follows:
\begin{itemize}
\item The paper describes the design and implementation of a new mobile \deleted{group}  
application \added{for group event scheduling} and its supporting system: OutWithFriendz provides smooth functions for group hosts to 
easily create an invitation and invite other users to join. All group members can 
suggest their \replaced{preferred meeting locations and date/time, }{ideal locations, meeting dates} and vote for them. The host finalizes meeting 
location and time based on the voting results.

\item The OutWithFriendz system represents the first field-based study of
\replaced{the group event scheduling and decision-making process }{event organization behavior}
in the context of a deployed mobile application with widespread geographic
usage. This has \replaced{allowed us }{provides the possibility} to collect precise
user \replaced{trace }{traces} data.

\item Using the data collected from a field study of this novel system,
we discovered a series of
factors, such as
mobility, host preference, user preference, and social voting influence that
are significant in group \added{event planning and} decision
making processes.
A correlation analysis of these factors is also performed. Our study offers
new insights for group hosts and members to improve their real-life
event organization.
\end{itemize}
The rest of this paper is organized as follows: After discussing the related
works, we introduce our system design and data collection. Next we present our
data analysis and correlation analysis in more detail.
Finally, we \added{summarize the important results, highlight the key findings,}
discuss \replaced{their potential usage }{the limitations}, and conclude this work.

\section{Related Work}
\label{sec:related}

In this section, we discuss works that are most relevant to ours.
We divide them into three categories.

\deleted{Group behavior analysis has been an area of active research in recent
years.  Li et al. designed a temporal-spatial method for group detection,
locating and tracking~\cite{li2016temporal}. Sen et al. designed a group
monitoring system for urban spaces~\cite{sen2014grumon}, and Jayarajah et al.
studied how users' mobility patterns change when they are within a
group~\cite{jayarajah2015need}.
The work by Jamil et al. proposed a hybrid sensing system for monitoring crowd
dynamics in large-scale events such as large religious
gatherings~\cite{jamil2015hybrid}. Brown et al.
investigated the differences between individual and group behavior with respect
to physical location~\cite{brown2014group}. However, previous works have limited
understanding of how groups make decisions and the influence of the group
host and participants.  Our system allows users to easily express their
preferences through suggesting and voting for meeting locations and times.
Analyzing the decision making process data collected within our system can
provide insights that enable groups to better organize events.}

Group event organization has been studied in a number of recent works. Yu et al.
proposed a Credit Distribution-User Influence Preference algorithm to recommend
potential participants to the group host~\cite{yu2015should}. Zhang et al. and
Pramanik et al. built models to predict event success in
Meetup~\cite{zhang2016predicting, pramanik2016predicting}. The work by Du et al.
collected a series of contextual factors to predict individuals' activity
attendance~\cite{du2014predicting}. However, all of these works collected their
datasets from large public social websites such as Meetup and Douban.
In these services, the group host makes the event decision on their
own.  The meeting location and time are settled when the invitation is created.
An interactive group \replaced{event scheduling }{decision making} process is missing in these services.

Social influence occurs when an individual's emotions, opinions, or behaviors
are affected by others. This phenomena has been observed in many domains. For
instance, Goyal et al. proposed an influence maximization method based on a
historical user action log~\cite{goyal2011data, goyal2010learning}. Li et al.
further incorporated friend and foe relations in social networks for influence
maximization~\cite{li2013influence}.
Using a Meetup dataset, Zhang et al. demonstrate the significant impact of group
leaders in making event decisions~\cite{zhang2016predicting}. In the Doodle
voting application, Zou et al. find that in open polls, the voting decisions by later
respondents are highly influenced by early and nearby
respondents~\cite{zou2015strategic}. In this paper, we analyze the difference
between the impact of the group host and participants on group \replaced{event planning }{decision
making}, as well as the behavior of early and late-coming voters, \added{and how early 
voters affect late voters,} which to the
best of our knowledge has not been previously studied.

\deleted{User mobility analysis has been an area of active research in recent years. In previous work,
human mobility patterns are analyzed to identify taxi
trips~\cite{ganti2013inferring}, to predict user's future
movement~\cite{baumann2015population}, and to monitor whether individuals are
affected by depressive mood disorders~\cite{canzian2015trajectories}. No
existing work investigates the influence of human mobility on group event
attendance.  Our study based on the dataset collected by our mobile application
is valuable for understanding this effect. Our extensive analysis will
show that user mobility prior to the group activity has significant impact
on the user's availability on the group event planing process.}

\added{Our work is also generally related to group behavior analysis. 
Sen et al. designed a group
monitoring system for urban spaces~\cite{sen2014grumon}, and Jayarajah et al.
studied how users' mobility patterns change when they are within a
group~\cite{jayarajah2015need}. Lampinen et al. investigated how users deal with group co-presence
to prevent conflictive situations~\cite{lampinen2009all}. There are also works studying group colocated interactions using mobile 
devices~\cite{fischer2013understanding, lucero2015mobile}.
The work by Brown et al.
investigated the differences between individual and group behavior with respect
to physical locations~\cite{brown2014group}. However, previous works have limited
understanding of how groups schedule and make decisions for events, or the influence of the group
host and participants.  Our system allows users to easily express their
preferences by suggesting and voting for meeting locations and time.
By analyzing group event planning process using data collected from our system,  
we can gain some useful insights for better group event organization.}

\section{System Design}
\label{sec:methodology} 

In this section, we describe the design, architecture, and implementation of the
OutWithFriendz system.  We also present \replaced{a walk-through example to 
illustrate the user workflow of our app for group event scheduling }{the typical user workflow through a
series of use cases}.

\subsection{System Architecture}
\begin{figure}
\centering
\includegraphics[width=0.9\linewidth]{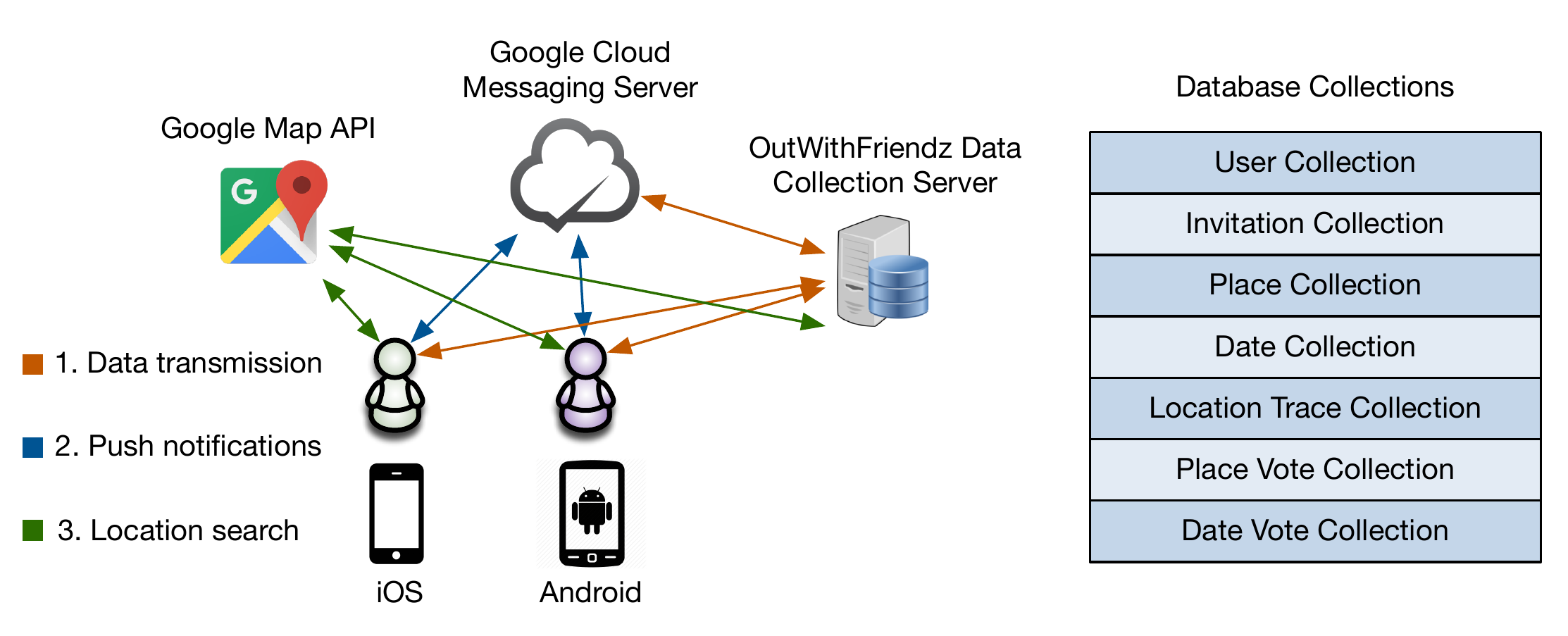}
\caption{The architecture of the OutWithFriendz system.}
\label{fig:architecture}
\end{figure}

In order to understand group user behavior at scale, we designed versions of the
OutWithFriendz mobile client for both the iPhone and Android platforms, allowing
us to collect data on the scale of hundreds of users and user-generated
invitations. This mobile system provides opportunity for us to track group
event organization behavior on a regular, ongoing basis. We support the two most
popular mobile platforms, iPhone and Android, allowing us to accommodate groups composed of
users of both platforms.
Our system consists of three major components: the mobile client, the data
collection server, and the Google Cloud Messaging (GCM)
server~\cite{developers2014google}. We also call the Google Maps API to retrieve
location search results.  Figure~\ref{fig:architecture} shows the overall
architecture of our OutWithFriendz system.

The implementation of the OutWithFriendz system requires careful engineering to
handle the communication and data synchronization between clients. Our
OutWithFriendz server is implemented as a Java Web application using the Spring
application framework~\cite{Spring}. All required functionality to the client is
exposed through the server's REST APIs. MongoDB is also used to store and manage
all data on the server~\cite{mongodb}. To push notifications between server and
clients, GCM services are used to handle all aspects of queueing of messages and
delivery to client applications running on both mobile platforms. In addition,
information about each location is obtained from Google Map
Services~\cite{googlemaps}, including the name, street address, and
latitude/longitude coordinates.

\begin{figure}
\centering
\begin{subfigure}{.33\textwidth}
  \centering
  \includegraphics[width=0.9\linewidth]{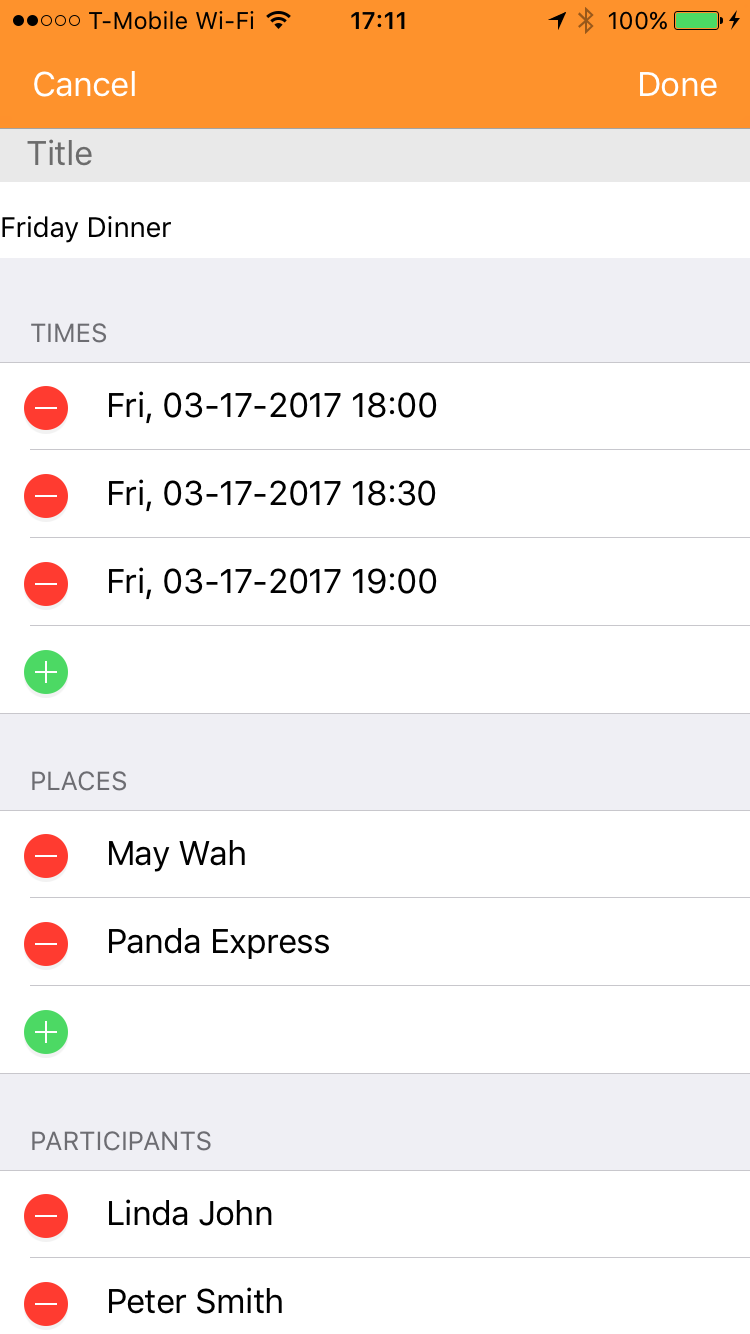}
  \caption{Create invitation.}
  \label{fig:screenshot_host}
\end{subfigure}%
\begin{subfigure}{.33\textwidth}
  \centering
  \includegraphics[width=0.9\linewidth]{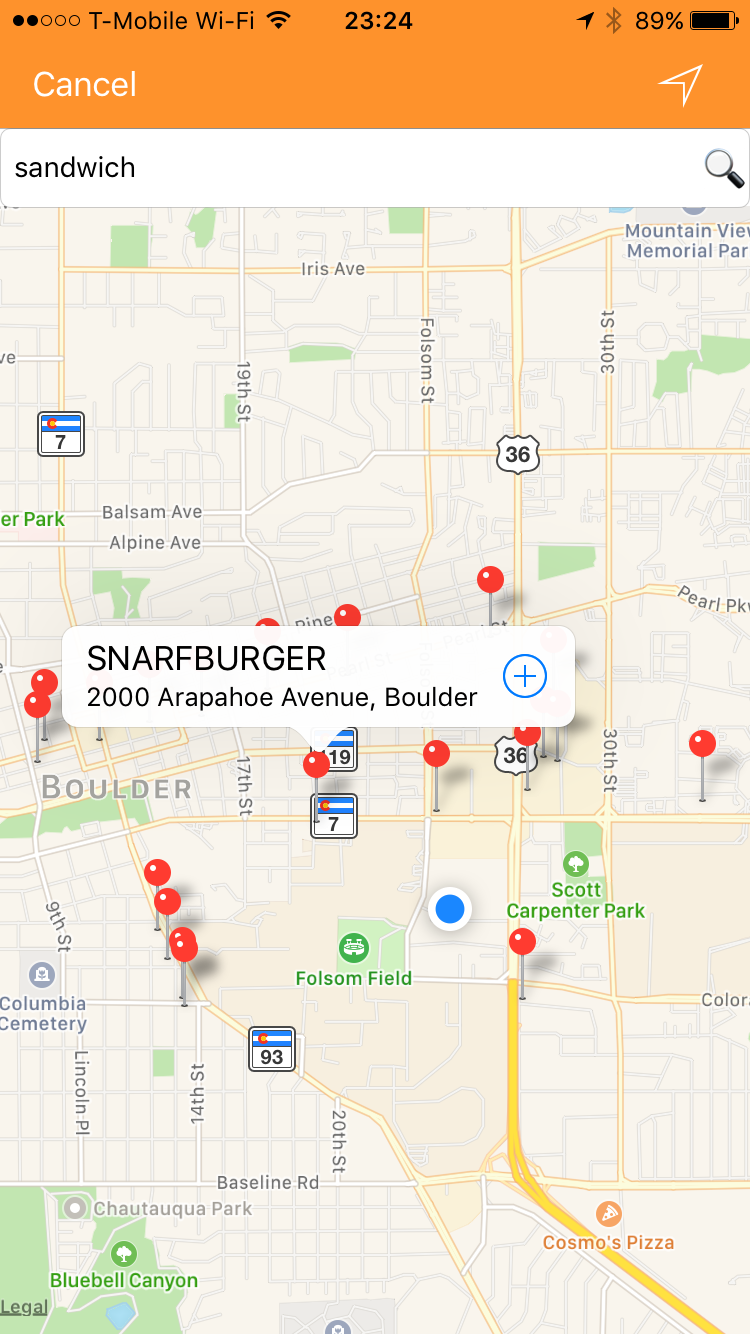}
  \caption{Add location through Google Map.}
  \label{fig:screenshot_map}
\end{subfigure}
\begin{subfigure}{.33\textwidth}
  \centering
  \includegraphics[width=0.9\linewidth]{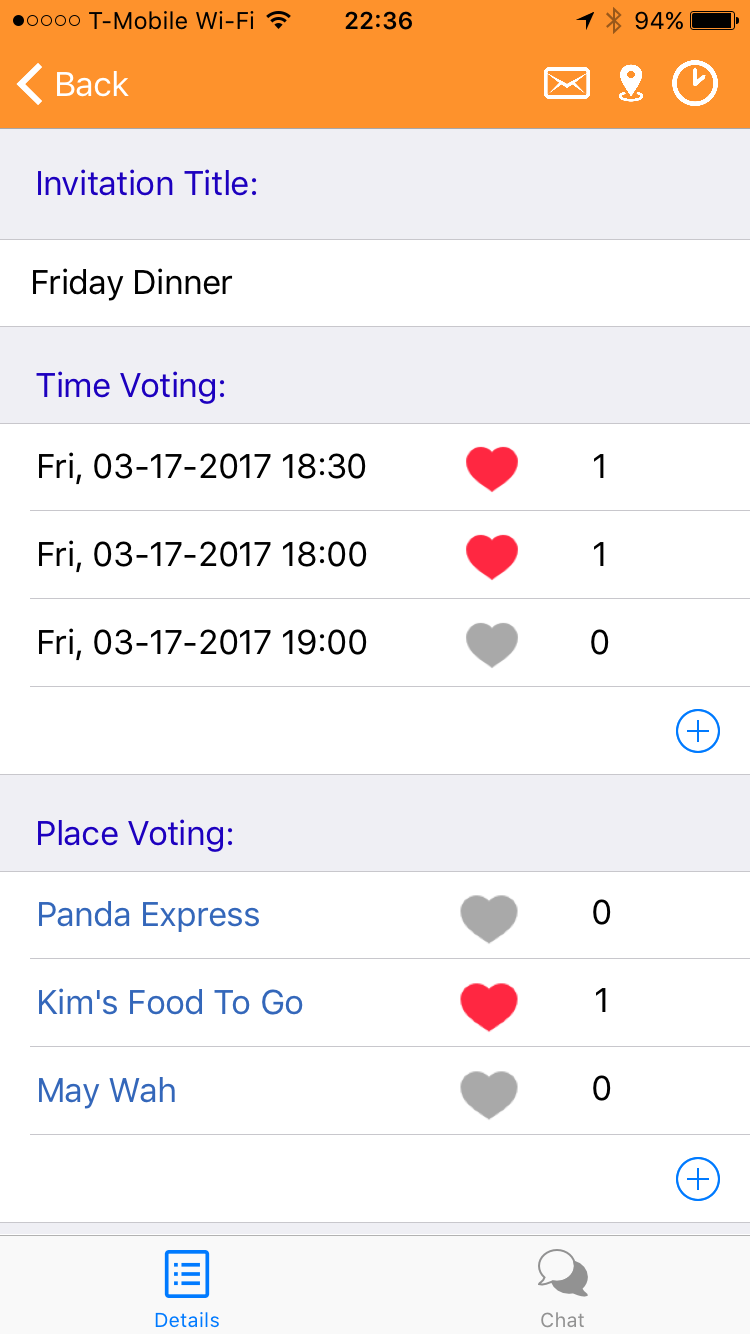}
  \caption{\added{Suggest and} vote for preferences.}
  \label{fig:screenshot_vote}
\end{subfigure}
\caption{Main workflow of the OutWithFriendz mobile application.}
\end{figure}

\subsection{UI Design Challenges}
In order to enable a natural group decision-making workflow, we continuously
streamlined the UI and workflow of the OutWithFriends app based on feedback
collected from user studies. We started with an initial usage survey
before releasing the app to the market. During our survey, we hired seven
students on campus who have different academic backgrounds. They formed three
groups to use our app and provided useful feedbacks for improving UI design.
For example, these users suggested: (1) Adding a chat board to allow group
members to discuss their opinions; (2) Allowing users to edit the location title
and provide detailed information for each location; (3) Allowing users to link
suggested locations with the Google Places application; (4) Pushing notifications
if an invitation is created or modified; and (5) Replacing text buttons with
interactive icon buttons. Implementing this functionality helped us improve our
app to better support the real-life group \replaced{event scheduling }{decision making} process.
We also added and altered application functionality to improve usability, based
on many user suggestions received during application usage.
At the beginning of the study, we focused on dining events only. Later we came
to realize that the users would also like to use the OutWithFriends app for
generic group gatherings, such as going for a hike or watching a movie. To
support this functionality, we shifted from integrating with Foursquare API to the more
suitable Google Places API. We also changed the workflow for the voting process
to make it more flexible. Initially, invitation participants were required to
decide on the meeting time before starting the voting process for the location.
However, our users preferred to perform time voting and location voting
concurrently, which is more flexible. Next, we describe the main workflow of our app.

\subsection{\replaced{A Walk-through Example }{Use Cases}}
\label{sec:flow}

\begin{figure}
\centering
\begin{subfigure}{.33\textwidth}
  \centering
  \includegraphics[width=0.9\linewidth]{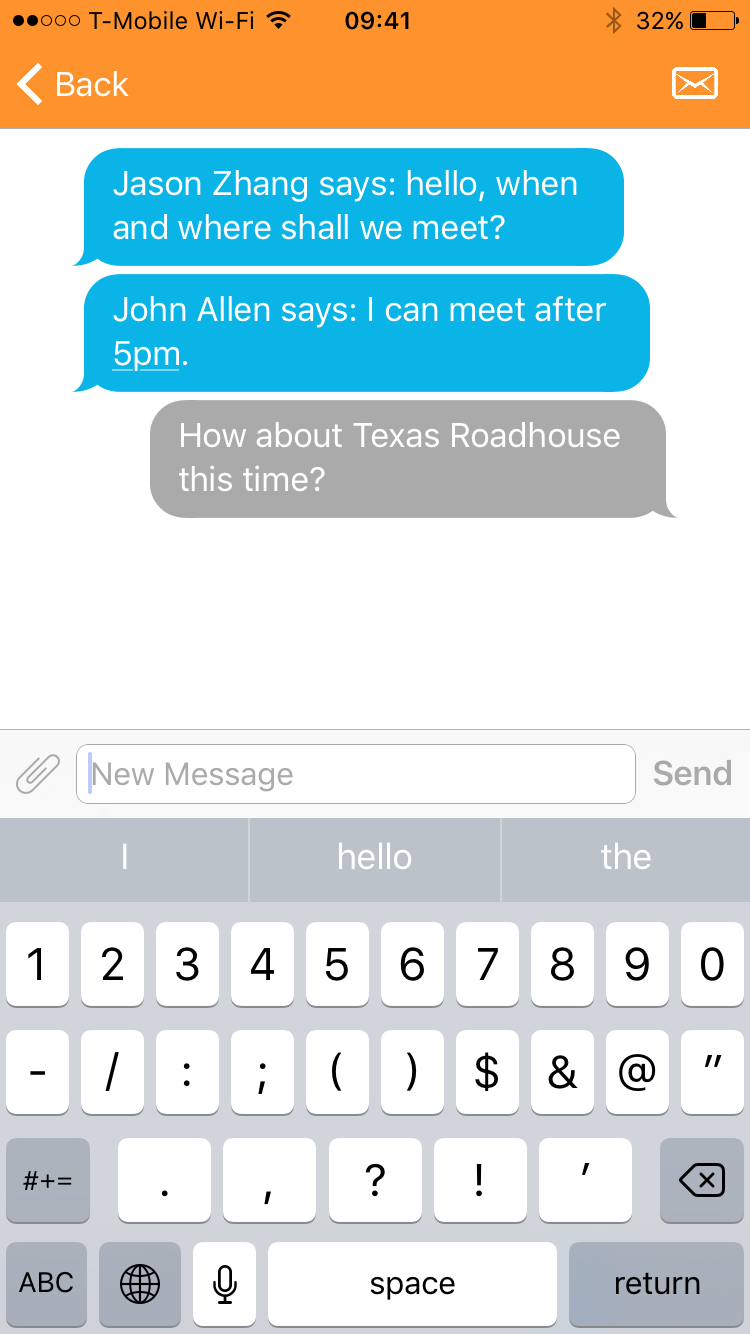}
  \caption{Chat screen.}
  \label{fig:screenshot_chat}
\end{subfigure}%
\begin{subfigure}{.33\textwidth}
  \centering
  \includegraphics[width=0.9\linewidth]{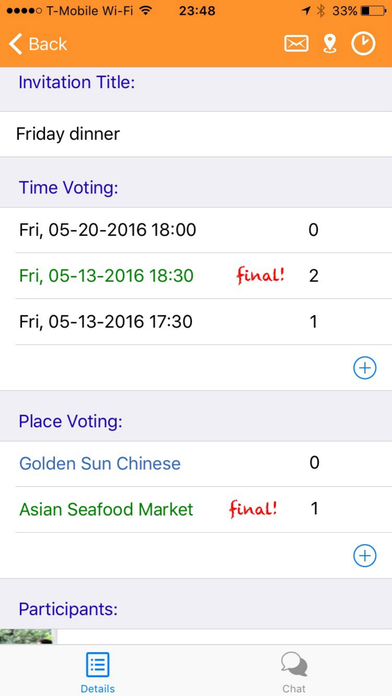}
  \caption{Invitation finalization screen.}
  \label{fig:screenshot_finalize}
\end{subfigure}
\begin{subfigure}{.33\textwidth}
  \centering
  \includegraphics[width=0.9\linewidth]{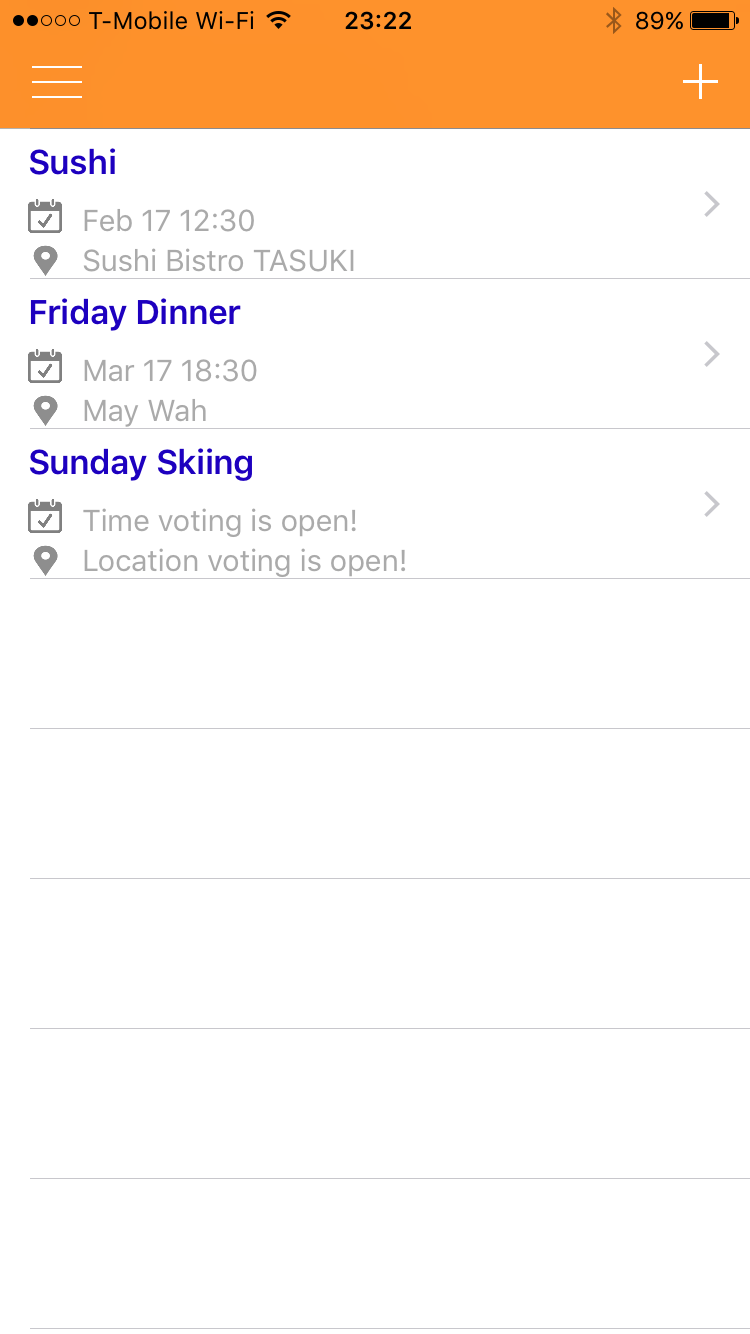}
  \caption{Invitation list screen.}
  \label{fig:screenshot_list}
\end{subfigure}
\caption{Main functions of the OutWithFriendz mobile application.}
\end{figure}

\added{To better understand the workflow of our mobile application, 
we provide a walk-through example of how the main functions are 
used for group event scheduling.}

\subsubsection{A host invites two friends to meet for dinner}
In this use case, we describe the actions a user would take to invite some friends to meet for 
dinner. Here we call this user the host. When creating 
a new event invitation, the host will go to the window shown in
Figure~\ref{fig:screenshot_host} and perform the following steps:
(1) create a title for the invitation, \added{such as Friday Dinner}; 
(2) specify one or more possible dates and times for the 
invitation; (3) suggest meeting locations using Google Map Services, as
shown in Figure~\ref{fig:screenshot_map}; and (4) add one or more friends that
want to be included as participants in the invitation. Finally, when the host is
satisfied with the invitation settings, she taps the ``send invitation'' icon, to send the invitation to all selected participants.
\added{She can also start voting for her own preferences right after the new invitation shows up on her screen.}

\subsubsection{A user receives an invitation to meet several friends for dinner}
First, the user receives a notification from the OutWithFriendz application
indicating that she has received a new invitation. The user can express her 
preferences by voting on one or more possible options for meeting dates and
locations, as shown in Figure~\ref{fig:screenshot_vote}. \added{One important feature of our system is that} 
the user may also add
new proposed dates/time or locations to the invitation. Once the user has added a new
option, it will be automatically made visible to all other participants. \added{Users are also allowed to change their suggestions 
and votes throughout the voting process.}
In the ``Chat'' tab shown in Figure~\ref{fig:screenshot_chat}, \replaced{the }{user}
is also able to send text messages to other group members \added{for discussion and better coordination 
of the scheduling process}.

\subsubsection{Host finalizes the invitation based on voting results}
The voting process continues until the host decides to finalize the meeting time 
and location. Only the host is permitted to finalize, which is shown in
Figure~\ref{fig:screenshot_finalize}. After the host has finalized the
invitation, each participant receives a notification regarding this action. 
\added{To support unforeseen changes, the host could still update the final decision 
after it is finalized.}
Each user's main screen will show a list of invitations that she has 
participated in, as shown in Figure~\ref{fig:screenshot_list}.

\section{Data Collection and General Characteristics} 
We first describe the dataset we collected using our OutWithFriendz system, then conduct a data distribution 
analysis to understand the key characteristics of our dataset.

\subsection{Data Collection}
\begin{figure}
\centering
\includegraphics[width=0.6\linewidth]{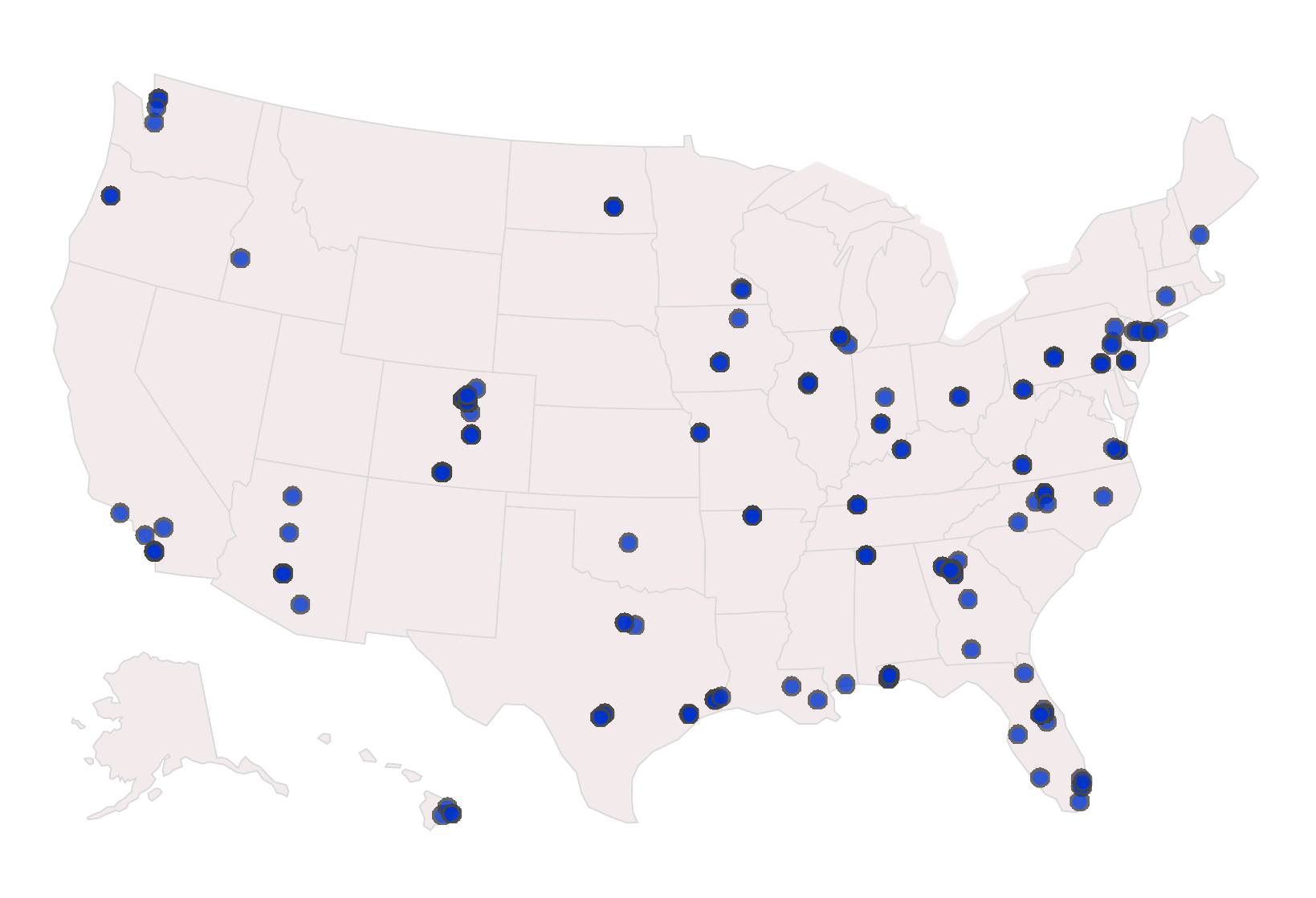}
\caption{The geographic distribution of all finalized locations across the US.}
\label{fig:map}
\end{figure}

We deployed our OutWithFriendz mobile application on the Google Play and Apple Store marketplaces.
To collect enough data for group dynamics analysis, we posted advertisements on 
Microworkers~\cite{microworkers} and Craigslist~\cite{craigslist}
for participants. For teaching these users how to use our app correctly, we also made an introductory 
video, which is included in our supplemental file: ``OutWithFriendzIntroductionVideo.mp4''. For each 
legitimate completed invitation, we paid the host of a group 20 dollars, with the provisions that: (1) The host 
and participants must live in the US; (2) The host should invite at least two other 
friends to the invitation using our app; 
(3) The group must demonstrate a full voting process; (4) The host must finalize the meeting time  and location
for the invitation; (5) Each participant would open their location services on their smartphone during the study
and allow us to track their mobility traces; (6) At least half of the group members attended the finalized 
event\footnote{\added{We added this requirement to prevent workers from creating fake invitations and making dishonest money. It would be interesting to analyze the low-attendance events, which we plan to investigate when we have more users and events.}}. From these 
two job post websites, we collected 246 legitimate invitations over a 5-month period from 432 users. In addition, 71 
students on our campus used the app without getting any payment, which contributed another
76 legitimate invitations. The whole data collection period spanned from January 2016 to May 2017. In total, 503 distinct users
of our OutWithFriendz application were identified, generating 322 legitimate invitations. \added{141 users have been the host of at least one event, and 72 hosts have created exactly one event each.} Please note that 
each user is allowed to create and join multiple invitations in this study. \added{Moreover, 11 groups (7 from paid users and 4 from students in our university) have used our mobile app frequently, with more than six legitimate invitations in our dataset.}
Figure~\ref{fig:map} shows the distribution of all suggested locations recorded in our server across the US. 
It indicates that our users are widespread in 34 
different states and 81 cities throughout the country.  

\begin{figure}
\centering
\includegraphics[width=1\linewidth]{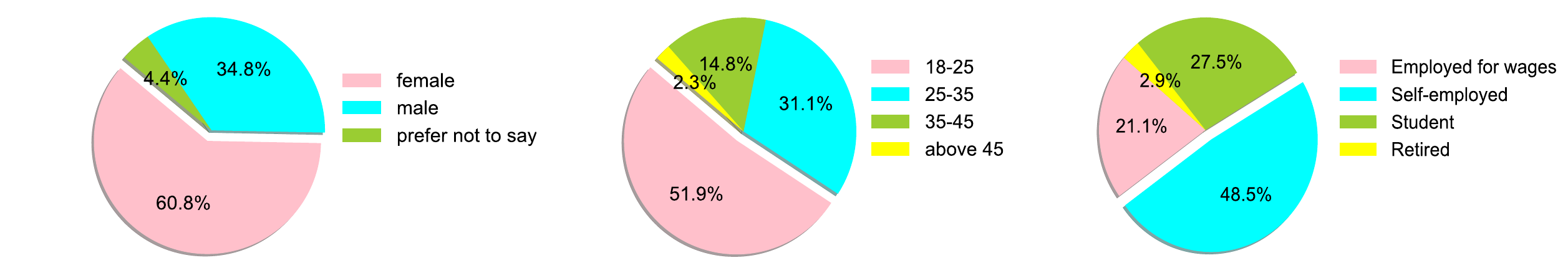}

\caption{User demographic information of OutWithFriendz Users (294 participants). (L) Gender distribution. (M) Age distribution. (R) Profession distribution}
\label{fig:demographic}
\end{figure}

\added{To better understand the demographics of our users, we have conducted an anonymized survey using
Google Form.  We contacted each user through his/her Facebook Page to complete the survey. 
In total, 294 of our users have completed the survey.
The results are shown in Figure~\ref{fig:demographic}. 
We can see that 60.8\% of the users who responded to the survey are
female and 48.5\% are self-employed. 
Students, including some from our campus and some from the Microworker website,
account for 27.5\% of our participants. In addition, most (83.0\%) of our users are young, aged 18-35. } 

\subsection{Location Trace Data}
\label{sec:trace}
In addition to collecting data about the event organization process, we also collected user mobility-related data.  The OutWithFriendz app posts GPS user location traces to the server
either every 5 minutes if the app is running in the background or every 30 seconds if the app is running
in the foreground. Before participation all of our users were required to provide informed IRB consent. They would  
turn on the location services on their smartphone during the test so we were able to collect the data.  
All location traces are anonymized and permission to use this anonymized data is provided when installing the 
app. The total amount of user traces data collected was about 1.1 GB. 

\subsection{Group Size Distribution}
\begin{figure}
\centering
\includegraphics[width=0.32\linewidth]{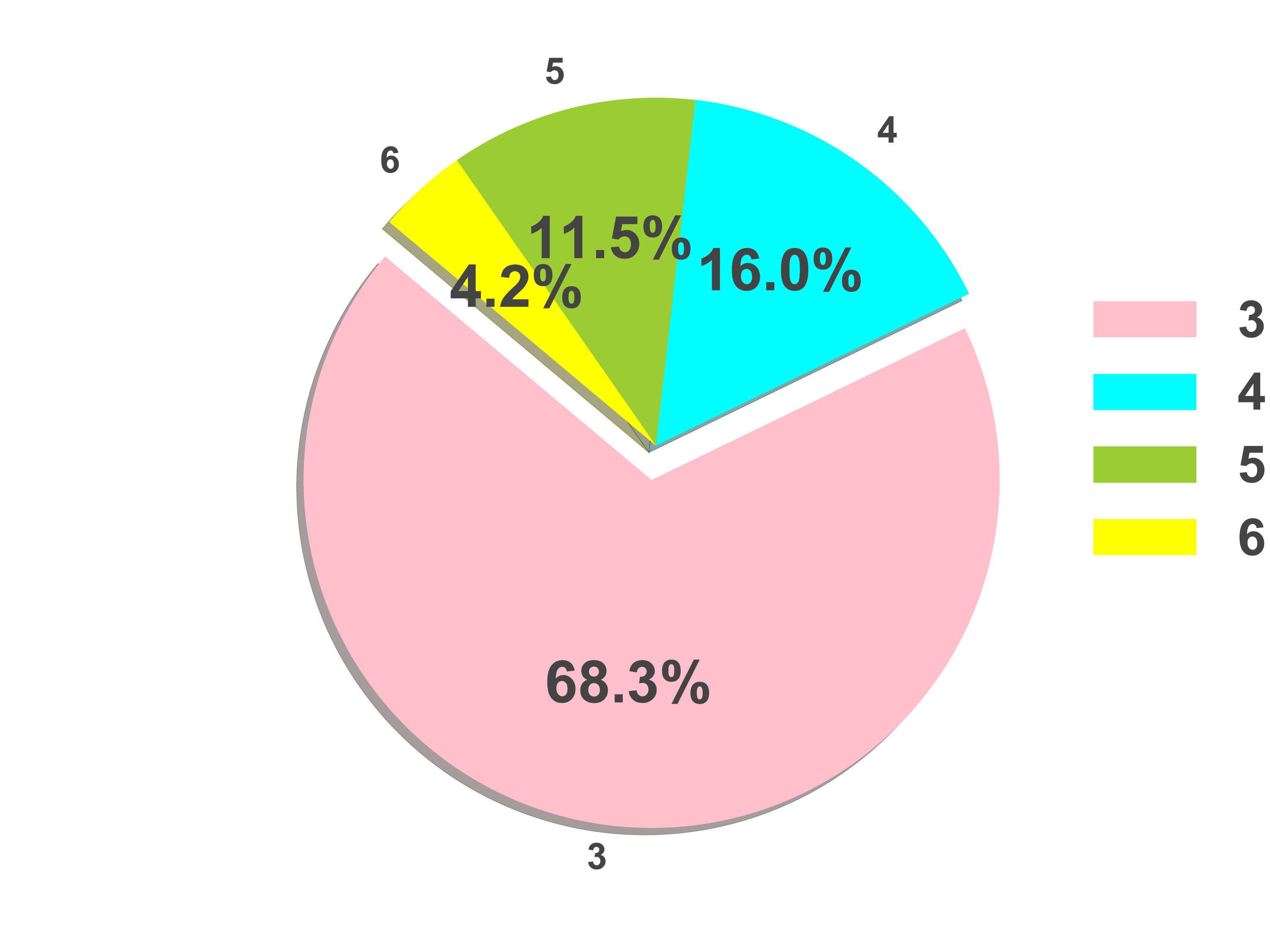}
\includegraphics[width=0.32\linewidth]{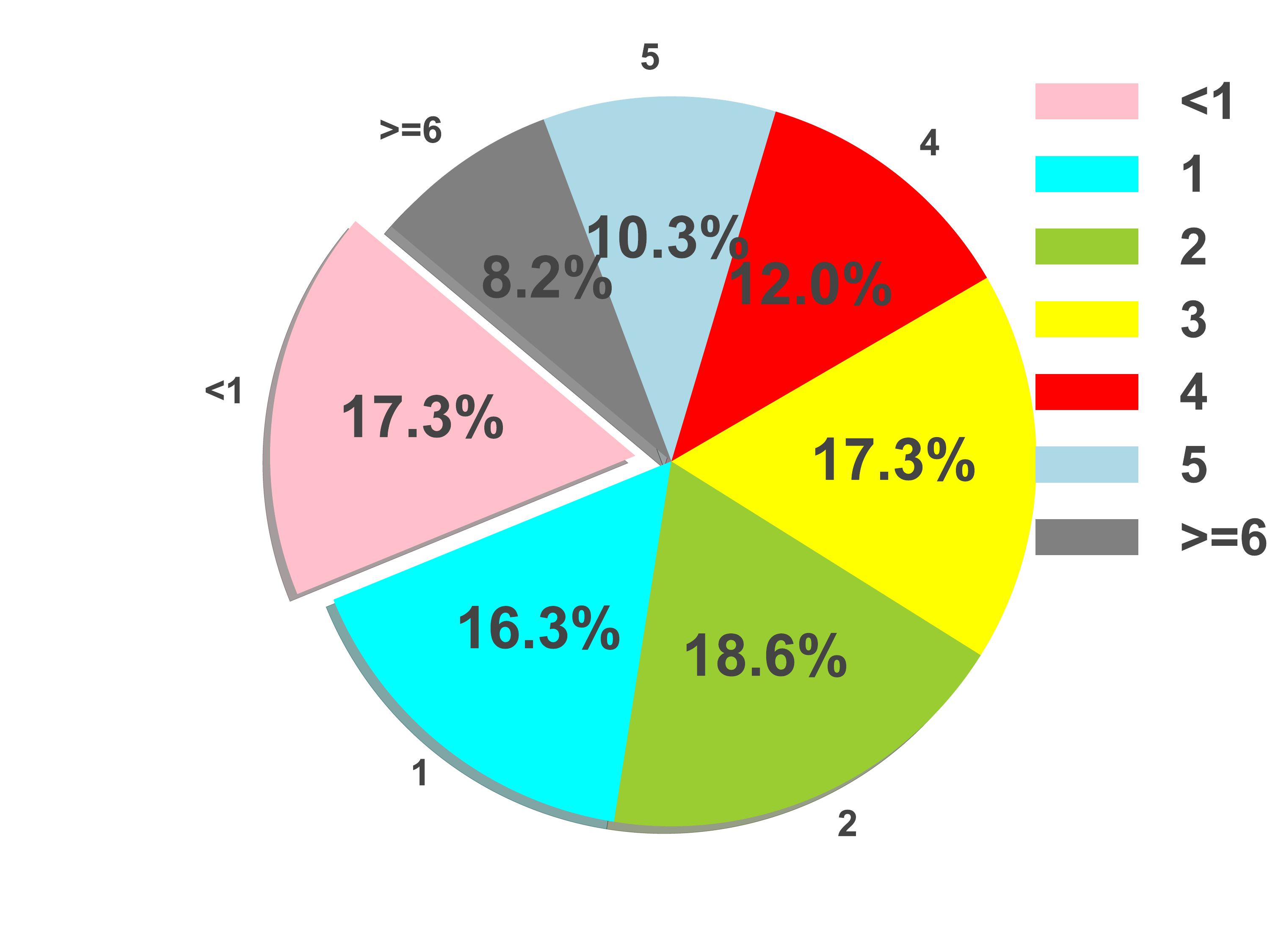}
\caption{The distribution of group size (left) and number of days to make final decision (right).}
\label{fig:groupsize}
\end{figure}

Figure~\ref{fig:groupsize} left summaries the distribution of group size in our OutWithFriendz dataset. \added{Here we define group size as the number of participants who finally stayed in the invitation. We do not count users who were removed from the invitation, either by themselves or by the host, because they did not  participate in the whole scheduling process and their votes were not shown after the removal.}  We observe that most of the groups in our study are small, with the large majority being groups of three.  We were pleased to see a significant fraction of groups with five (11.5\%) and six members (4.2\%) who were able to use the app concurrently.
Our work focused primarily on obtaining data for groups of three or more, which we feel represent many typical social group interactions of interest to us.  As a result, we did not focus on examining pairwise groups in this study.  The figure's trend lines suggest that if we had opened up our study to pairwise groups, then our data would have been overwhelmingly skewed toward pairwise groups.  However, now that we have obtained substantial initial data for larger groups, we plan to also explore the behavior of pairwise groups in our future works.

\subsection{Distribution of Days to Make Final Decision}
We are interested in the duration that it took for event organizers to make their final decisions. 
\replaced{As shown by the right figure in Figure~\ref{fig:groupsize}, the number of days to make the final decision is somewhat evenly distributed, and there is no dominant duration in this distribution. This is a bit  surprising or counter-intuitive, since we expected that there may be a more pronounced duration of decision-making within the first couple of days of creating an invitation. }{It is a little
surprising that we find the distribution of days to make the final decision to be somewhat evenly distributed,
which is illustrated in Figure~\ref{fig:groupsize} right.  No duration dominates this distribution, while we expected that there might be a more pronounced bunching of decision-making in the first couple of days.} However, there are also a substantial fraction of events that took four or more days to decide (about 30\%), indicating that a large fraction of hosts are taking a long time to decide. 
\added{This may be affected by the type of events and the amount of lead time. For daily meals, users can make a decision within thirty minutes while for some weekend activities, they will start planning it at the beginning of the week.}

\subsection{Voting Distribution of Individual Users}
\begin{figure}
\centering
\includegraphics[width=0.32\linewidth]{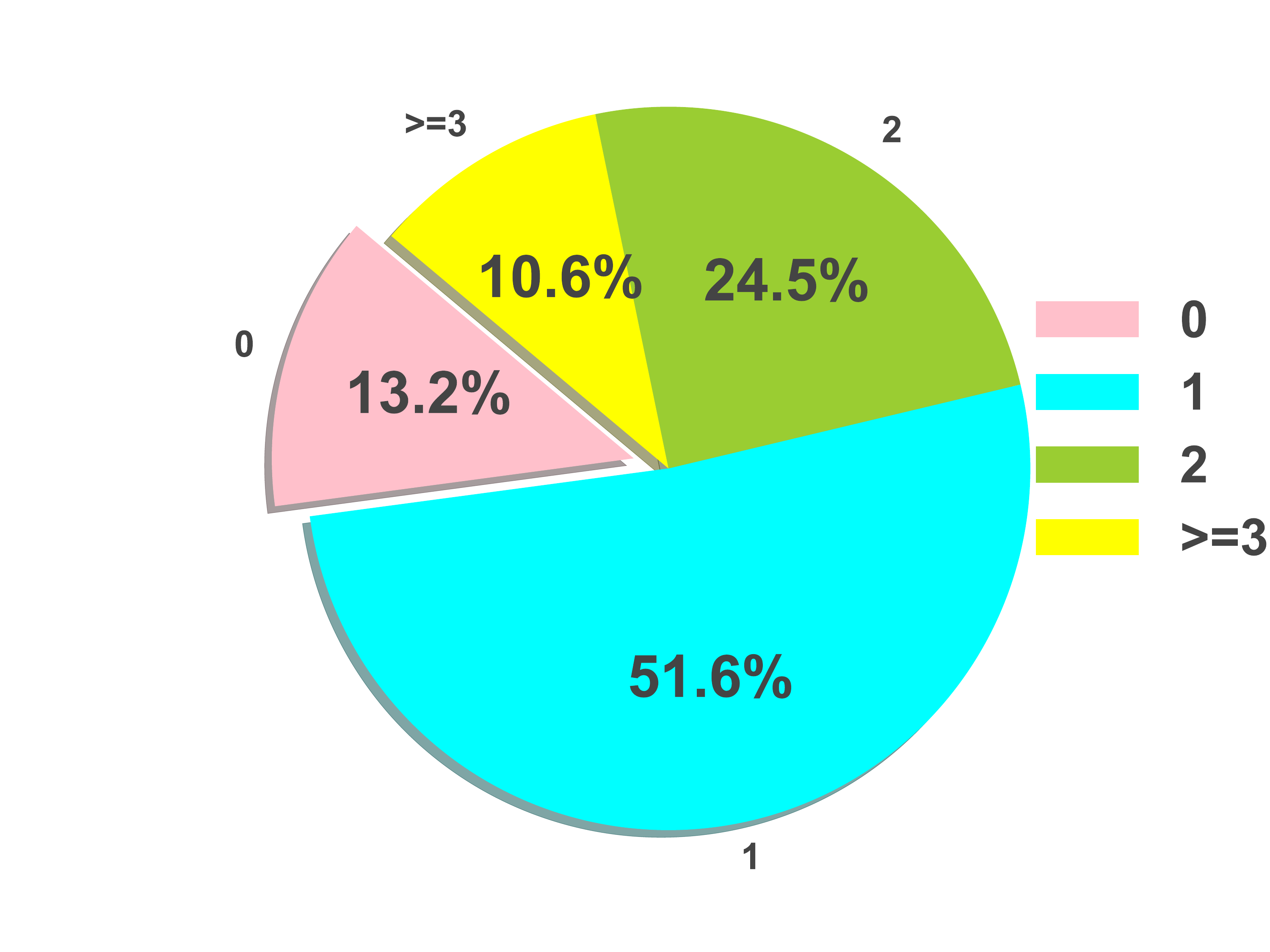}
\includegraphics[width=0.32\linewidth]{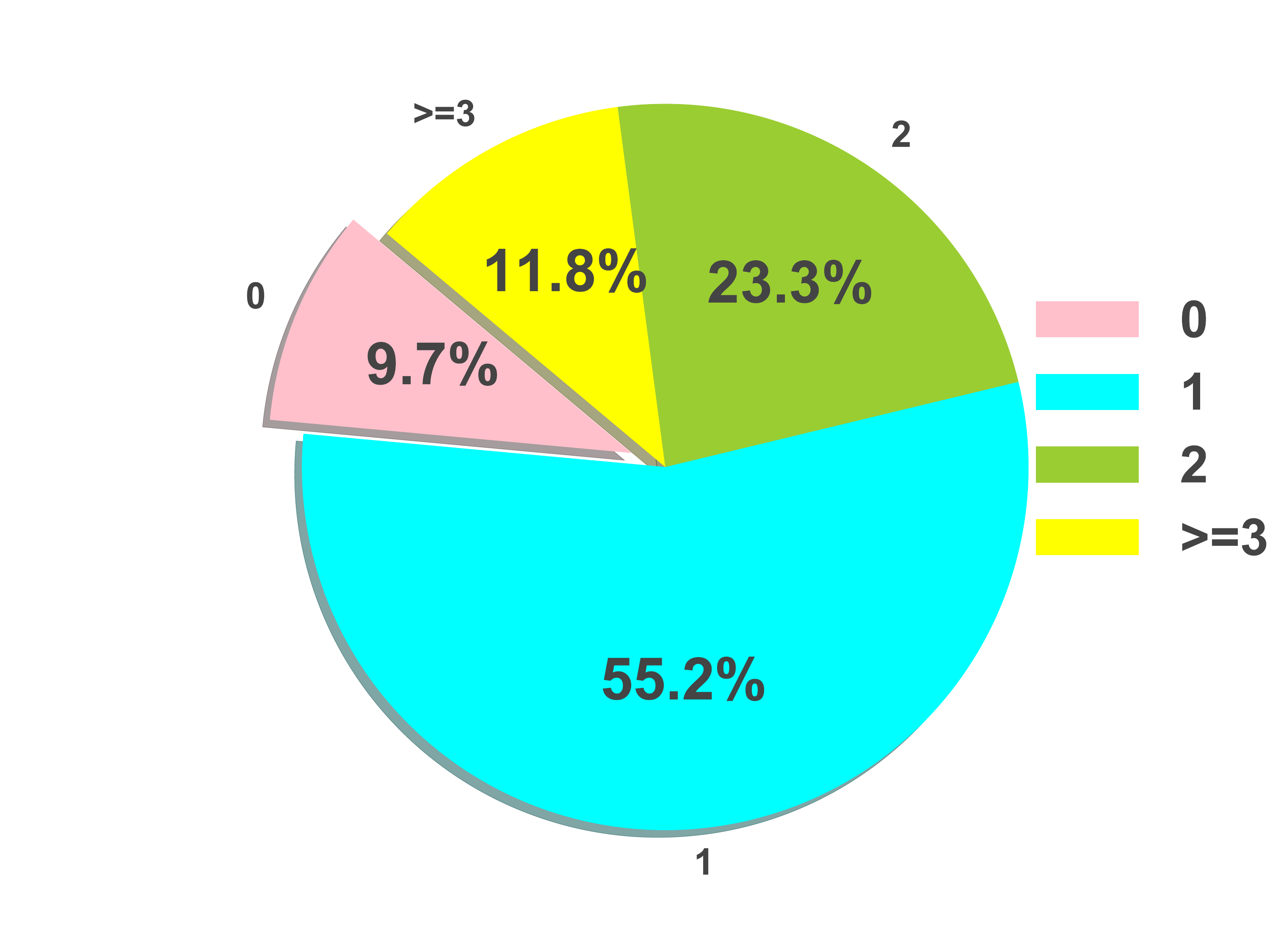}
\caption{The distribution of number of votes by a single user for event time (left) and place (right).}
\label{fig:votedis}
\end{figure}

Voting distribution is based on the number of votes made per individual user. The distributions for time 
and place voting are shown in Figure~\ref{fig:votedis}. The majority of users will vote for one 
option as far as the event time.  Similarly, the majority of users will vote for one option in terms of the place voting. In both processes, around 10\% didn't vote and 10\% voted for more than
2 options.  This voting behavior is analyzed further in later sections.

\subsection{Distribution of the Proportion of the Votes}

\begin{figure}
\centering
\includegraphics[width=0.7\linewidth]{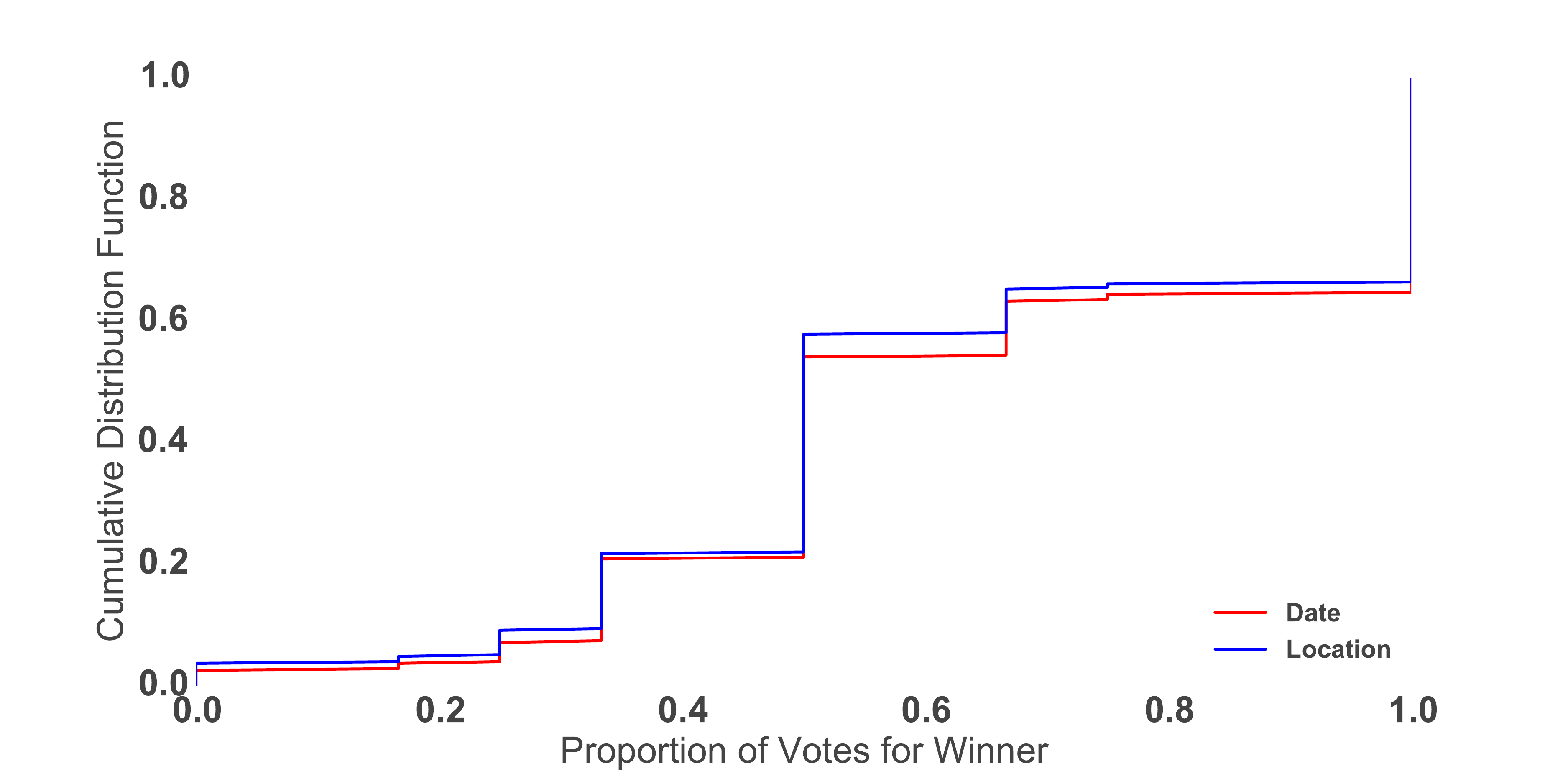}
\caption{The proportion of users who voted for final event time (red) and location (blue).}
\label{fig:winner}
\end{figure}

\replaced{We also analyze the proportion of users who voted for event time or location in the final decision.} 
{We conduct an analysis that refers to the proportion of votes the poll winner achieved versus the other contenders.}
The distribution is shown in Figure~\ref{fig:winner}. 
More than 70\% of the \replaced{final decisions }{winners} for both time and location
received majority votes to become the final choice. \added{This is understandable since groups tend to 
agree on the majority votes.} For the remaining 30\%, we observe some
very interesting behavior. In these polls, the final \replaced{decisions}{winners} did not receive the majority of the votes. In fact, in a small fraction of cases, there is a non-zero proportion of polls in which the \replaced{final decision}{winner} received no votes.  In these cases, the group host, who is the only one with the power to finalize the \replaced{event time and location }{place/date}, decided to override the \added{majority} voting results, either by personal fiat or possibly through a discussion with other group members that caused them to change their minds.

\subsection{Suggestion Distribution}
\begin{figure}
\centering
\includegraphics[width=0.32\linewidth]{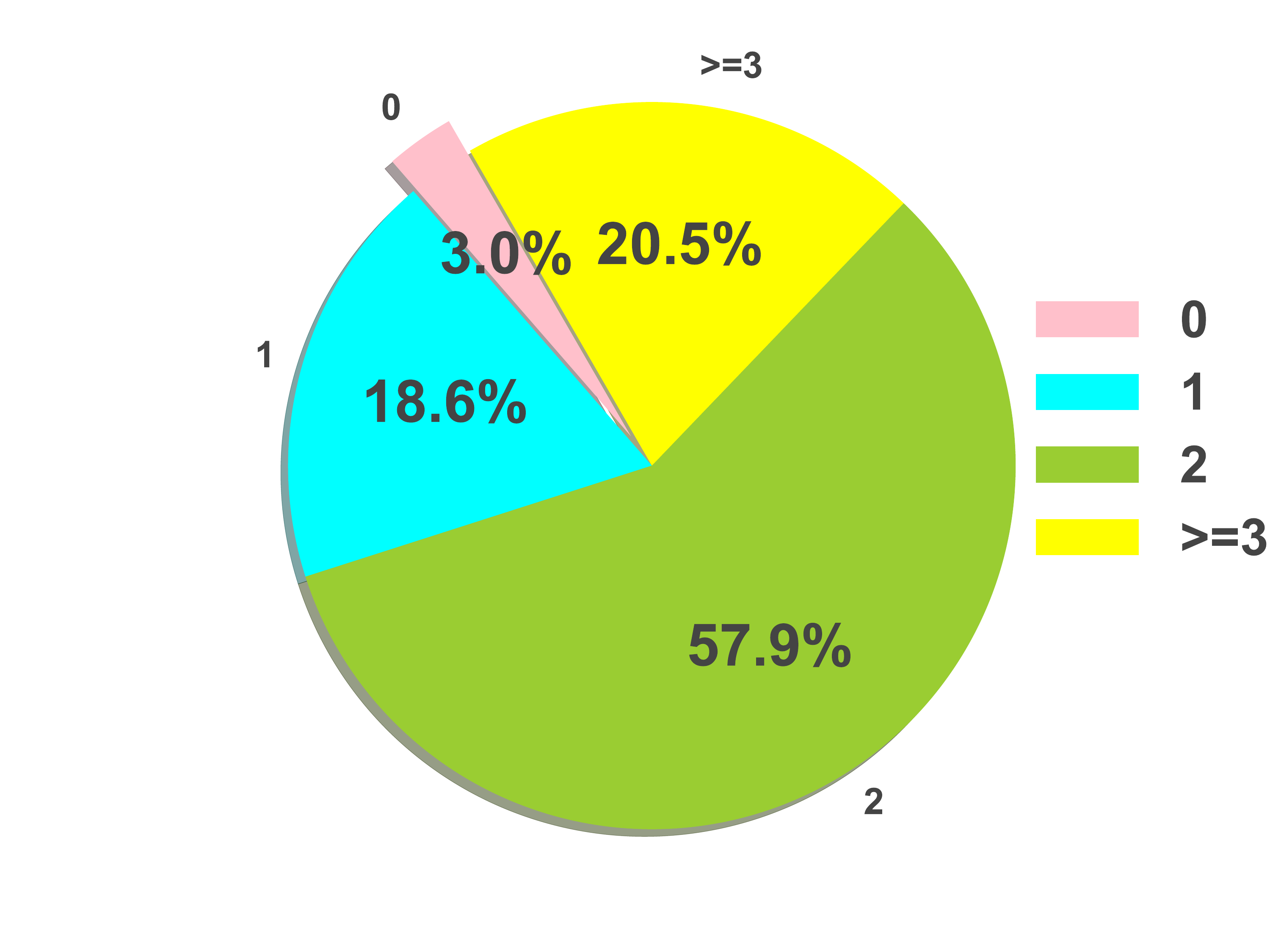}
\includegraphics[width=0.32\linewidth]{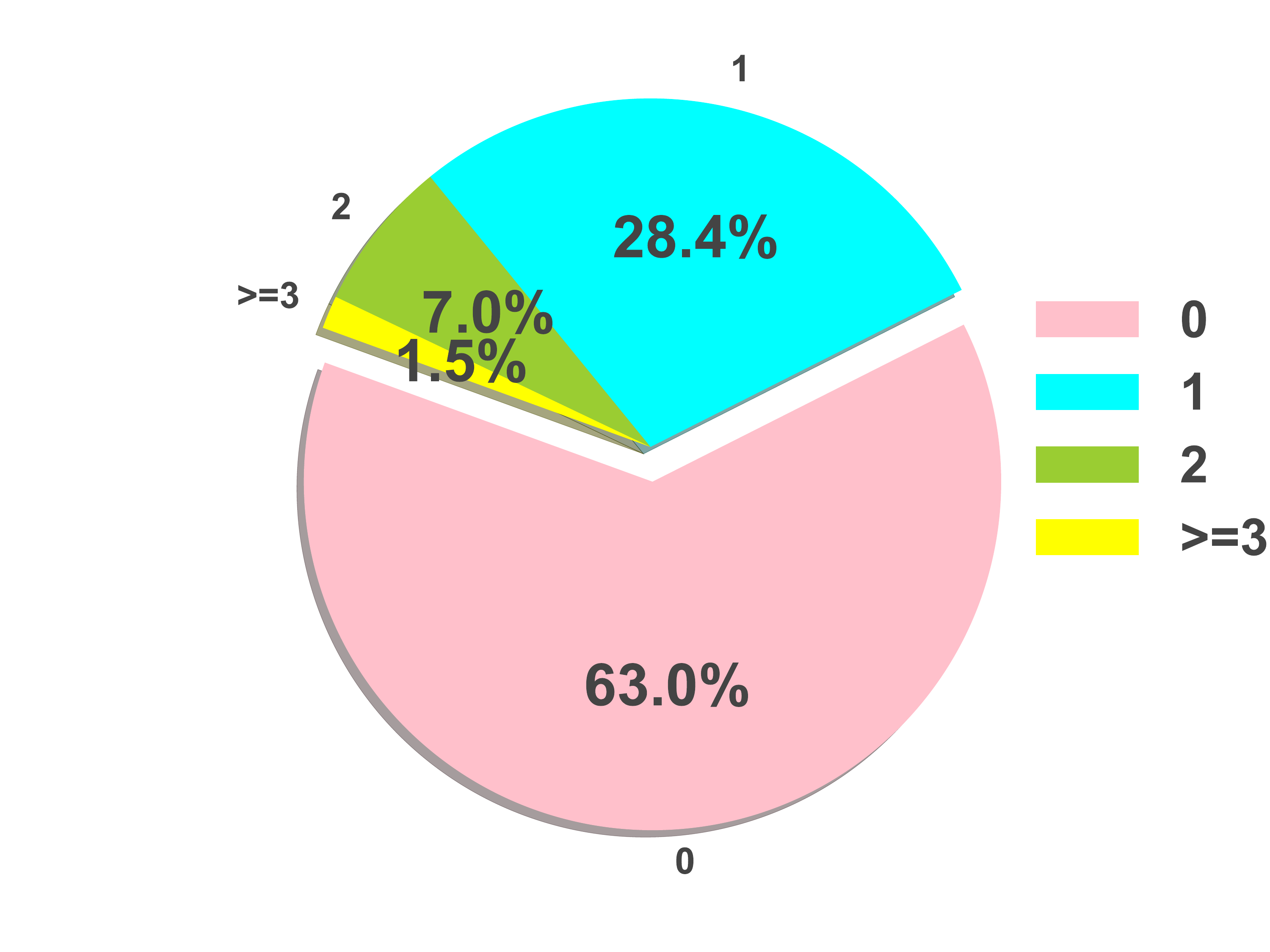}
\caption{The distribution of the number of suggestions made by group host (left) and other participants (right).}
\label{fig:suggest}
\end{figure}

OutWithFriendz app allows group participants not only to vote for their preferences, but also to suggest 
new options. Figure~\ref{fig:suggest} shows the suggestion distributions for host and participants.
Most hosts will suggest 2 or more options for the event. We also observe a small portion of hosts who provide no options of their own, and rely on other group members to provide suggestions.  For
participants, more than 60\% didn't make new suggestions. They just vote for the existing options. 
Some made one new suggestion while very few of them would make too many new suggestions. 
We will further compare the influence of group host and participants in our group decision section.

\deleted{We calculated the gender distribution of our OutWithFriendz users, which were 
about 42\% male and 58\% female. Please note that our app requires a user's 
Facebook account to log in, but it is not permitted by Facebook to get the accurate 
gender information using Facebook user ID. So here we estimate users' gender based
on manually reading their Facebook name and profiles. It is possible that some users' genders
are miscounted here but we expect the proportion to be rare. }

\subsection{\added{Metro vs Non-metro Areas}}

\added{Using the location trace data we collected from our users and the U.S. census data, we are able to identify locations frequently visited by our users. The technique we used will be introduced in detail in Section 5. Then we can project each user's home county using the frequently visited locations. According to the latest Rural-Urban Continuum Codes released in May 2013~\cite{rural}, every county is classified as a non-metro or metro area. In our dataset, 48\% the users live in metro areas and 52\% live in non-metro areas.} \footnote{\added{Please note our dataset contains 71 students who lived in Boulder doing this study. If we remove this student population, the proportion of metro and non-metro users would be 39\% and 61\%, all from crowdsourcing market users.}}

\subsection{Weather Factor}
An important external factor that can influence event organization relates to weather. Here we 
examine the impact of rain and temperature on our dataset. For this analysis, weather and
temperature information for each event was scraped from weathersource 
API~\cite{weathersource} at its location and starting time in our dataset. Note that we can only 
get hourly weather data from weathersource. If the starting time of one event is 19:35, in the 
analysis we use 19:00 weather data at the same day crawled from weathersource. Here we 
decide it is raining if the precipitation of an event's starting time is above 0. On snowy days, usually 
the precipitation will also be above 0, which we classify as rainy in our analysis.

\begin{figure}
\centering
\includegraphics[width=0.9\linewidth]{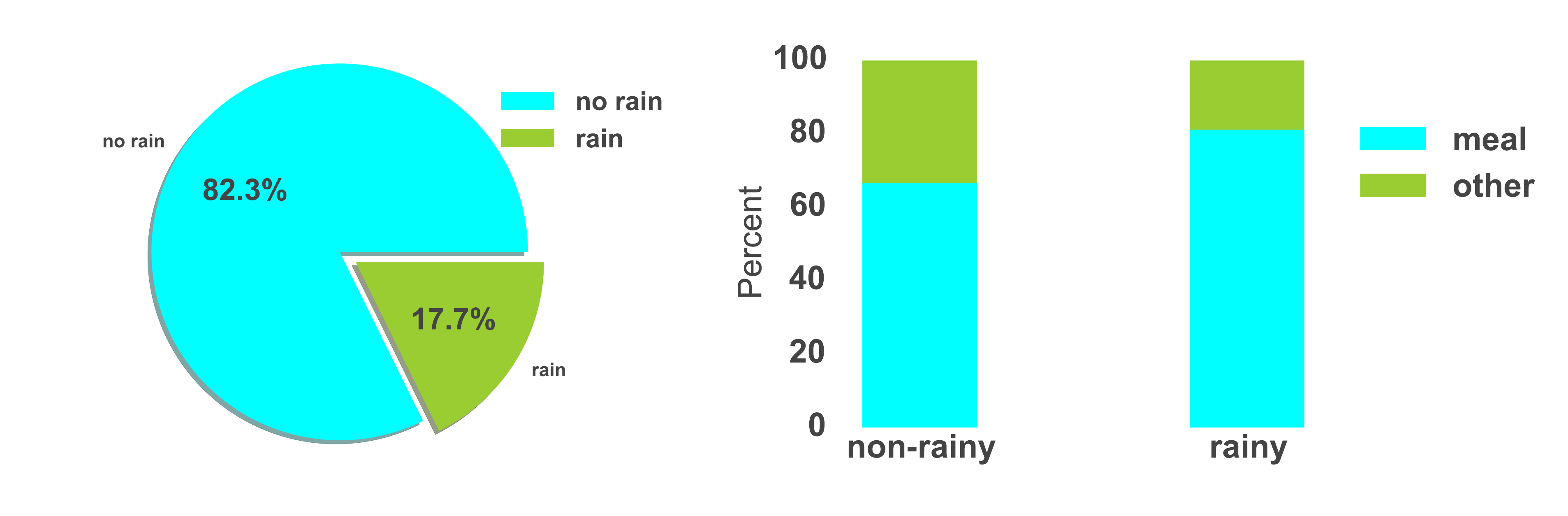}

\caption{(L) The distribution of events on rainy days vs. non-rainy days; (R) The 
distribution of meal events and other events on non-rainy days and rainy days.}
\label{fig:weather}
\end{figure}

Figure~\ref{fig:weather} shows the distribution of events that happened in 
rainy weather or not. 82.3\% of events organized in our app
occur in non-rainy weather. There are two reasonable explanations:
(1) In many places around the country non-rainy days happen more often than rainy
days; (2) Bad weather would have negative influence on real event attendance.
Looking deeper, bad weather appears to affect the types of events that are organized.
In OutWithFriendz, any place can be added to the Google Map as an option for voting, and need 
not be confined to a restaurant only.  For example, people have used the app to organize events such
as outdoor hiking and going to the movies.  We divide all events into two category types: 
meal events and other events. Meal events refer to people hanging out for lunch or
dinner, which is the majority event type in our dataset.  Other events include activities
that are not primarily dining, e.g. sporting and entertainment events.  In our study, we found 
that bad weather would have less impact on meal events compare with other types of events.
Figure~\ref{fig:weather} show that 66.7\% of the events belong to meal events on non-rainy 
weather while this number goes up to 81.1\% on rainy days.

\section{Group Decision Analysis} 
\label{sec:analysis}  
The analysis in this section examines the impact of a number of factors on 
group decision. First, we define some concepts and notations that will be used throughout this section. In the 
following analysis, we only use completed invitations in our dataset. A ``group
decision'' refers to the information submitted by the host after an event has
occurred, including the final group consensus rating.

In addition, for each event $e$, we define $T_e$ as the set of suggested meeting
times and $L_e$ as the set of suggested locations. For each participant $i$ and
option $o$ of event $e$, we let $V(i, o)$ be an indicator function which
indicates whether $i$ voted for option $o$:
\begin{equation}
V(i, o) = \left\{ \begin{array}{rcl}
1 & \mbox{Participant $i$ voted for option $o$} \\ 0 & \mbox{Otherwise} 
\end{array}\right.
\end{equation}

Then we define user's available time and location options as:
\begin{equation}
\label{time_avail}
\added{\text{user $i$'s time availability for event $e$}} = \frac{1}{|T_e|}\sum_{o = 1}^{T_e}V(i, o)
\end{equation}
\begin{equation}
\label{loc_avail}
\added{\text{user $i$'s location availability for event $e$}} = \frac{1}{|L_e|}\sum_{o = 1}^{L_e}V(i, o)
\end{equation}
\deleted{where $A_t(i, e)$ and $A_l(i, e)$ refers to user $i$'s time availability and location availability for event $e$.}

\newtheorem{observation}{Observation}
\subsection{Impact of User Mobility on Group Decision}
\label{sec:impact-user-mobility}
\begin{figure}
\centering
\begin{subfigure}{.5\textwidth}
  \centering
  \includegraphics[width=0.9\linewidth]{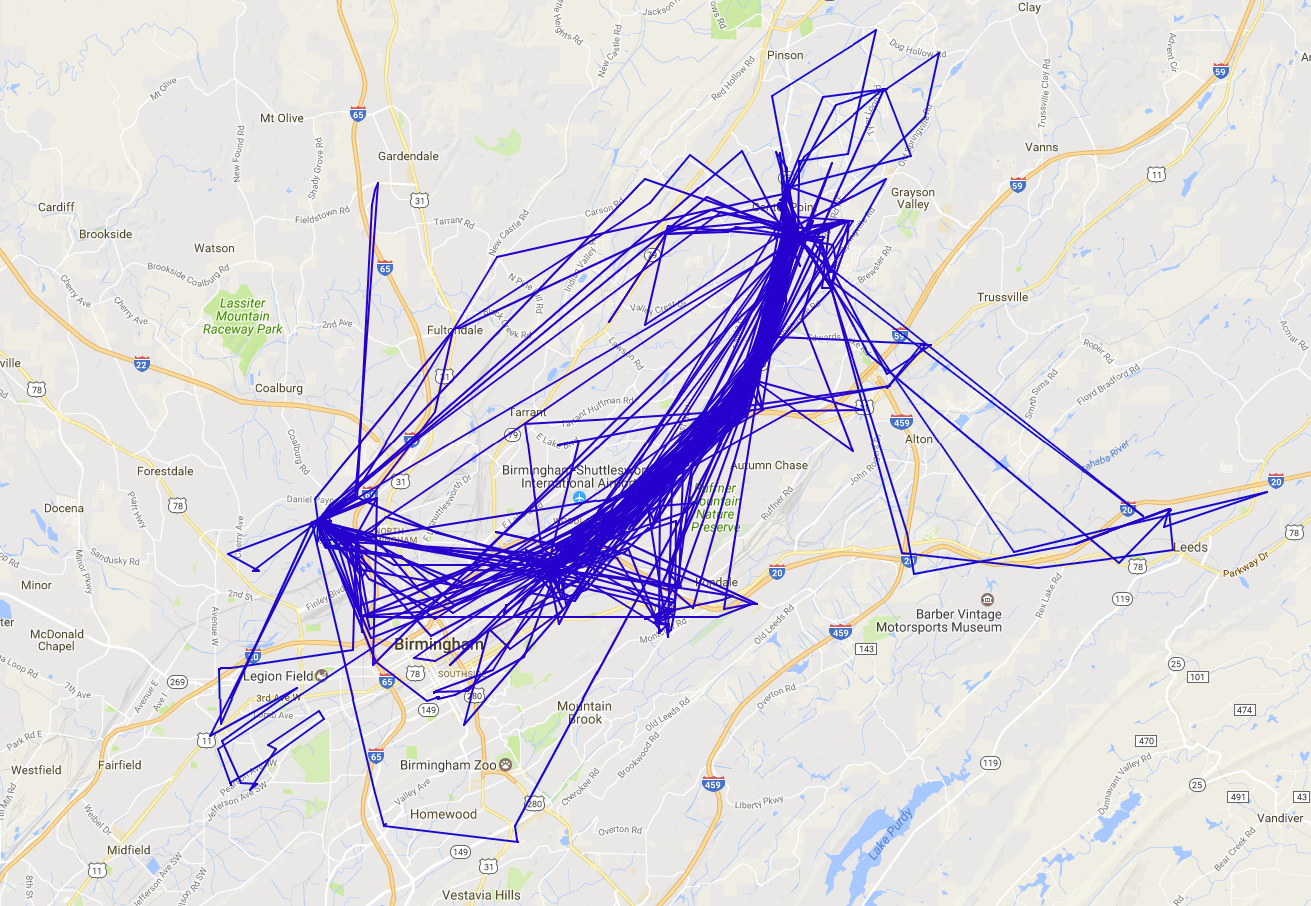}
  \label{fig:cluster1}
\end{subfigure}%
\begin{subfigure}{.5\textwidth}
  \centering
  \includegraphics[width=0.9\linewidth]{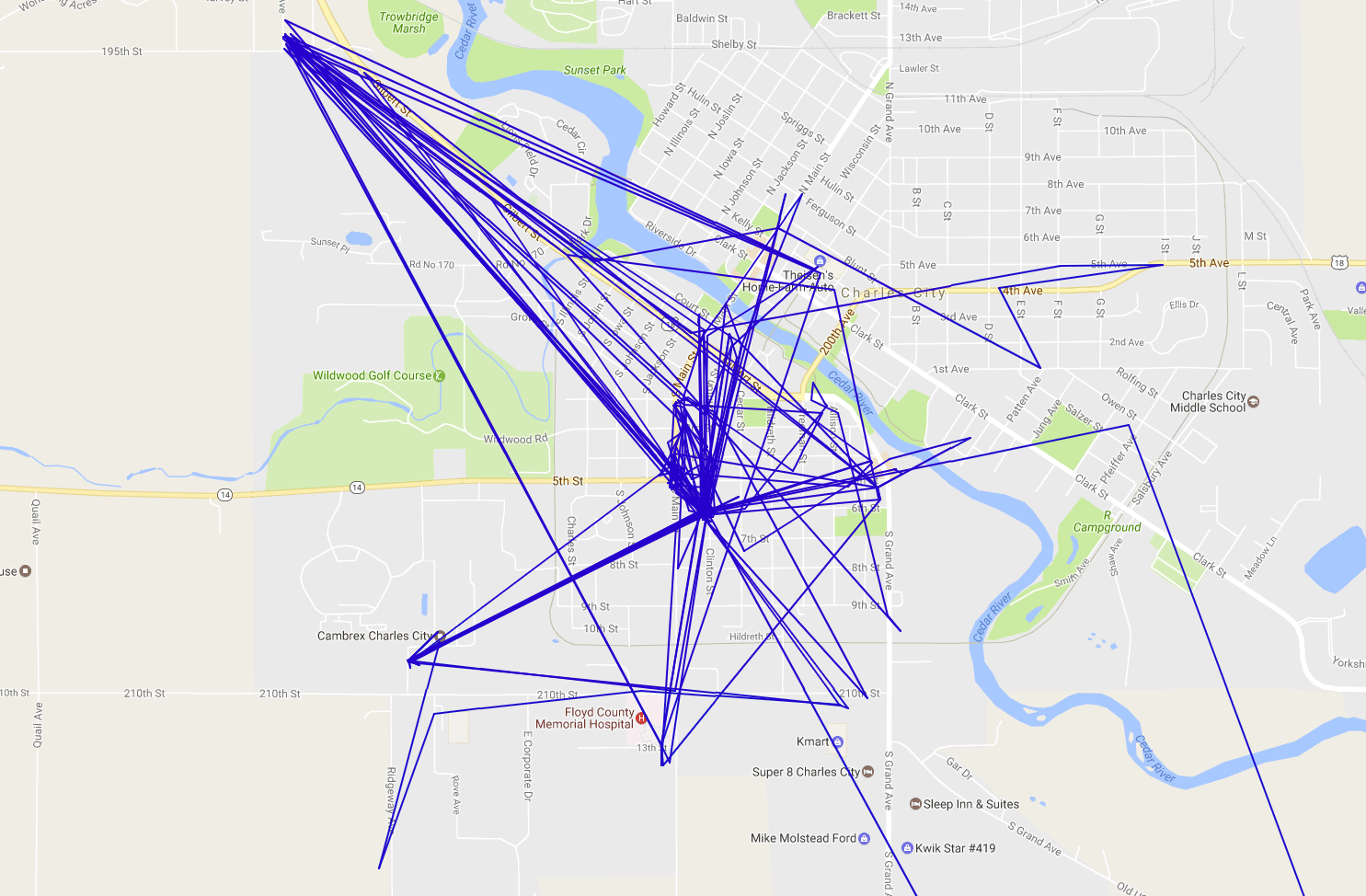}
  \label{fig:cluster2}
\end{subfigure}
\caption{User traces of two OutWithFriendz users. }
\label{fig:cluster}
\end{figure}

Using an individual user's location trace data, we are able to analyze
statistical properties of individual mobility. One way to consider a movement is
to calculate the distance between two consecutive location trace points in our
dataset.  This will result in the detection of many very short movements, such
as from one office to another in the same building. However, due to the location
services limitation in today's mobile phones, these short movements cannot be
traced precisely. Figure~\ref{fig:cluster} shows two examples of user traces
recorded in our dataset. There are natural clusters in these location traces
which appear to correspond to locations frequently visited by users, such as
work, school, and home.
To eliminate these very short movements and extract long movements, we implement
an algorithm introduced by Ye et al~\cite{ye2009mining}, which was originally
designed for GPS data. Assume that each individual's location trace points
detected by mobiles devices are ordered by timestamp $L = \{l_1, l_2, l_3,...,
l_n\}$. We identify two types of movements.
\textit{Type 1} refers to the short movements of a user within a building. In
\textit{Type 2}, the user will travel from one area to another with a
significant travel distance larger then $r$, for some period of time. In our
experiments, $r$ is set to 0.12 miles (200 meters) and the period threshold is
set to 30 minutes, as suggested by~\cite{ye2009mining}. To extract all the
\textit{Type 2} movements and eliminate \textit{Type 1} movements, we
iteratively seek spatial regions where the user remains for more than 30 minutes
and all the tracked points within this spatial region lie within 0.12 miles.
Then the location points in this spatial region are fused together by
calculating the centroid of these points. The centroid point is considered as a
\textbf{stationary point} for the spatial region.

\deleted{
When the spatial regions are detected, we calculate movement distance as the
travel distance from one region to another. We ignore all the trivial movements
that occur within the same spatial region. Figure~\ref{fig:moves} shows the
distribution of single movement distances calculated by our proposed algorithm.
The average movement distance is 5.35 miles on weekends and 4.12 miles on
weekdays. This pattern seems reasonable, as people will be more likely to travel
long distances for activities on weekends. The distribution of number of
movements per day is shown in Figure~\ref{fig:trips}. The average movement count
on weekdays is 5.3 and 5.6 during weekends. We also observe that
the proportion of very low movements and very high movements is higher on
weekends than weekdays. One possible explanation could be that people prefer
either rest on weekends, or to go out for activities on weekends,
which results in either low-movement or high-movement days.
We also compare our results with a US National Household Travel Survey conducted in
2009~\cite{santos2011summary}, which reports 4.2 movements during weekdays and
3.9 during weekends. Our numbers seem to be higher than the study in this
survey. Our hypothesis is that some of our OutWithFriendz users may close their
location services when they are not planning to use the app, in order to
prolong batter life.
These days are more likely to be their inactive days. Taking this into
consideration, it is not surprising that the average movement detected by our
app is slightly higher than reported by the survey.}

We now examine the impact of user mobility on group behavior in OutWithFriendz. 
Here we define user mobility as the total travel distance traveled by a user in
the 48-hour period preceding an invitation. Our assumption before was that users
who traveled longer distances will be more exhausted, and thus less likely to
have significant voting availability. However, our analysis refutes this
conjecture:
\begin{observation}
Users with higher mobility are more active in attending social events.
\end{observation}
We use the Pearson correlation coefficient~\cite{lawrence1989concordance} to
calculate the relationship between user mobility and voting availability.
Table~\ref{tab:mobility} shows that the correlation of user mobility with both
date and location voting availability is positive, and the results are
significant ($p<0.001$).
These results indicate that highly mobile users are more available for event
attendance.  There are two reasonable explanations for this phenomenon:
\begin{itemize}
\item Previous studies have shown that users who travel by car, bus, and foot
in daily life differ substantially in their value of time, in both revealed-preference and stated-preference 
surveys~\cite{liu1997assessment, elgar2004car}. In our OutWithFriendz dataset, the users who 
travel long distances may travel by car. This increases their likelihood of
attending events far away from their frequented spots. 
\item Users who have higher mobility are more likely to be active event
attenders. They are used to meeting with friends after school or work, which
results in longer travel distances. Conversely, office workers who sit
at their desks during the day have little mobility detected, but may still be
tired after work and less likely to travel.
\end{itemize}

\begin{table}[]
\centering
\caption{The correlation of user mobility and voting availability}
\label{tab:mobility}
\begin{tabular}{|l|c|c|}
\hline
                                                         & Pearson correlation coefficient & p-value  \\ \hline
The correlation of user mobility and date voting availability. & 0.276                             & 7.12e-05 \\ \hline
The correlation of user mobility and location voting availability. & 0.281                             & 2.92e-06 \\ \hline
\end{tabular}
\end{table}

\begin{observation}
\added{Group mobility has a positive correlation with an area's development degree.}
\end{observation}
\added{Given the spatial regions that are detected, we are interested in investigating whether there exists any pattern between a group's mobility and an area's degree of development. 
Our hypothesis is that groups living in metro areas have higher mobility than groups living in non-metro areas,  since metro group members may be more spread out in big cities and generate longer travel distances.
To perform this analysis, we downloaded the 2016 U.S. area development degree data from the U.S. Census Bureau~\cite{uscensus}.
Here we use population density and number of housing units to calculate the development degree of an area. For simplicity, we consider the location of each group event and that area's development degree. 
It is possible that group members live in a city but traveled to a rural area for the event. But this is rare in our dataset. 
Table~\ref{tab:population} shows the relationship between group's total travel distance and the corresponding county's population density and housing units. The Pearson correlation coefficient for these two parameters are positive with p-values that are smaller than 0.05. }

\begin{table}[]
\centering
\caption{The correlation of group mobility and area's development degree. }
\label{tab:population}
\begin{tabular}{|l|l|l|}
\hline
                   & Pearson Correlation Coefficient & p-value \\ \hline
Population density & 0.1834                          & 0.013   \\ \hline
Housing unites     & 0.1572                          & 0.018   \\ \hline
\end{tabular}
\end{table}

\subsection{Impact of Individual Preference on Group Decisions}
\label{sec:impact-individual-preference}
To discover underlying factors that may lead users to vote for specific event
options, we first focus our analysis on individual users. A social event is typically characterized by two major factors: event time
and location. Using the OutWithFriendz dataset we have collected, we first analyze the
travel distance between event suggested locations and each participant's closest location cluster, 
with the requirement that this cluster must contain a point with a timestamp that occurs within 2 hours
before or after the finalized time for the invitation. The 
suggested location options are further divided into two categories: the location
options with votes and location options without votes. Based on the results, we
make the observation:

\begin{figure}
\centering
\includegraphics[width=0.7\linewidth]{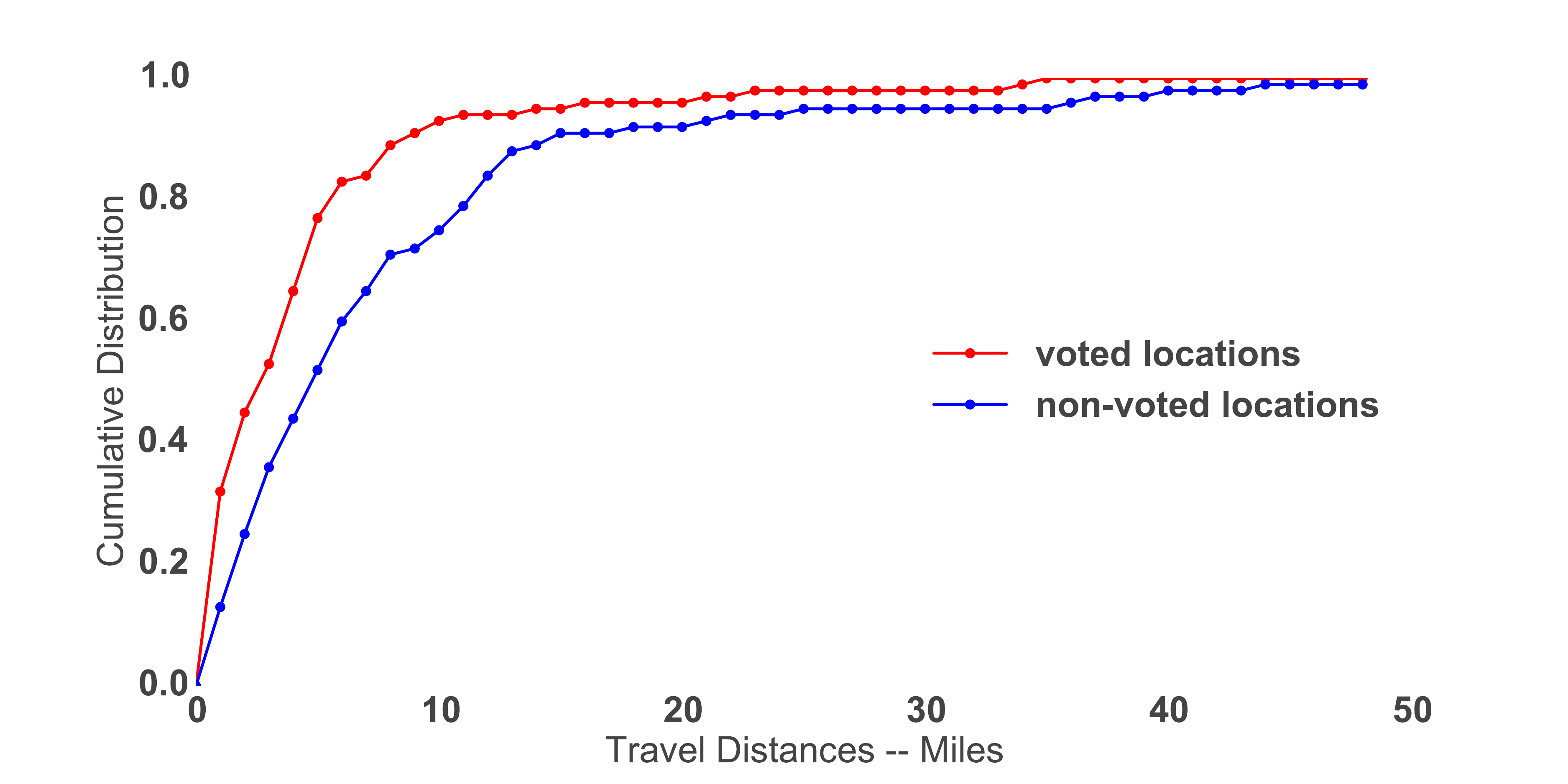}
\caption{The cumulative distribution of travel distances among voted locations and non-voted locations for each participant.}
\label{fig:distance}
\end{figure}

\begin{observation}
Most users would like to vote for event locations near their frequented
locations.
\end{observation}
Figure~\ref{fig:distance} shows the cumulative distribution of travel distances
among locations voted for and not voted for by each invitation participant.
The average travel distance for voted locations is 4.19 miles while for
non-voted locations is 7.53 miles. A Wilcoxon test found this to be a
significant difference $(z = -4.57, p<0.001)$, which indicates users have clear
preference to attend events near their frequented places. This is reasonable in
daily life. For example, we would intuitively expect that users would prefer to
go to dinner at restaurants that are close to their office or home.

\begin{figure}
\centering
\includegraphics[width=0.48\linewidth]{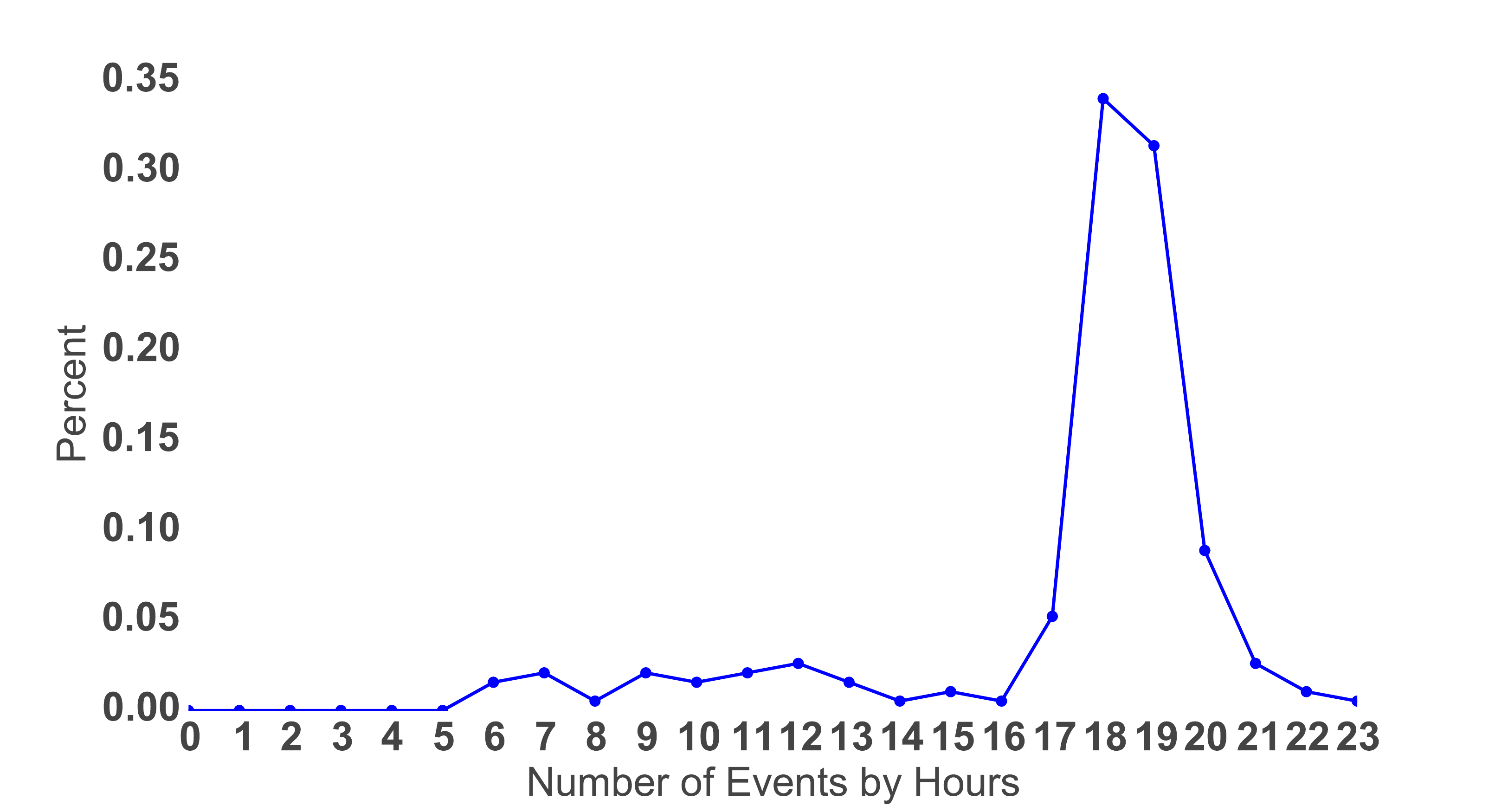}
\includegraphics[width=0.48\linewidth]{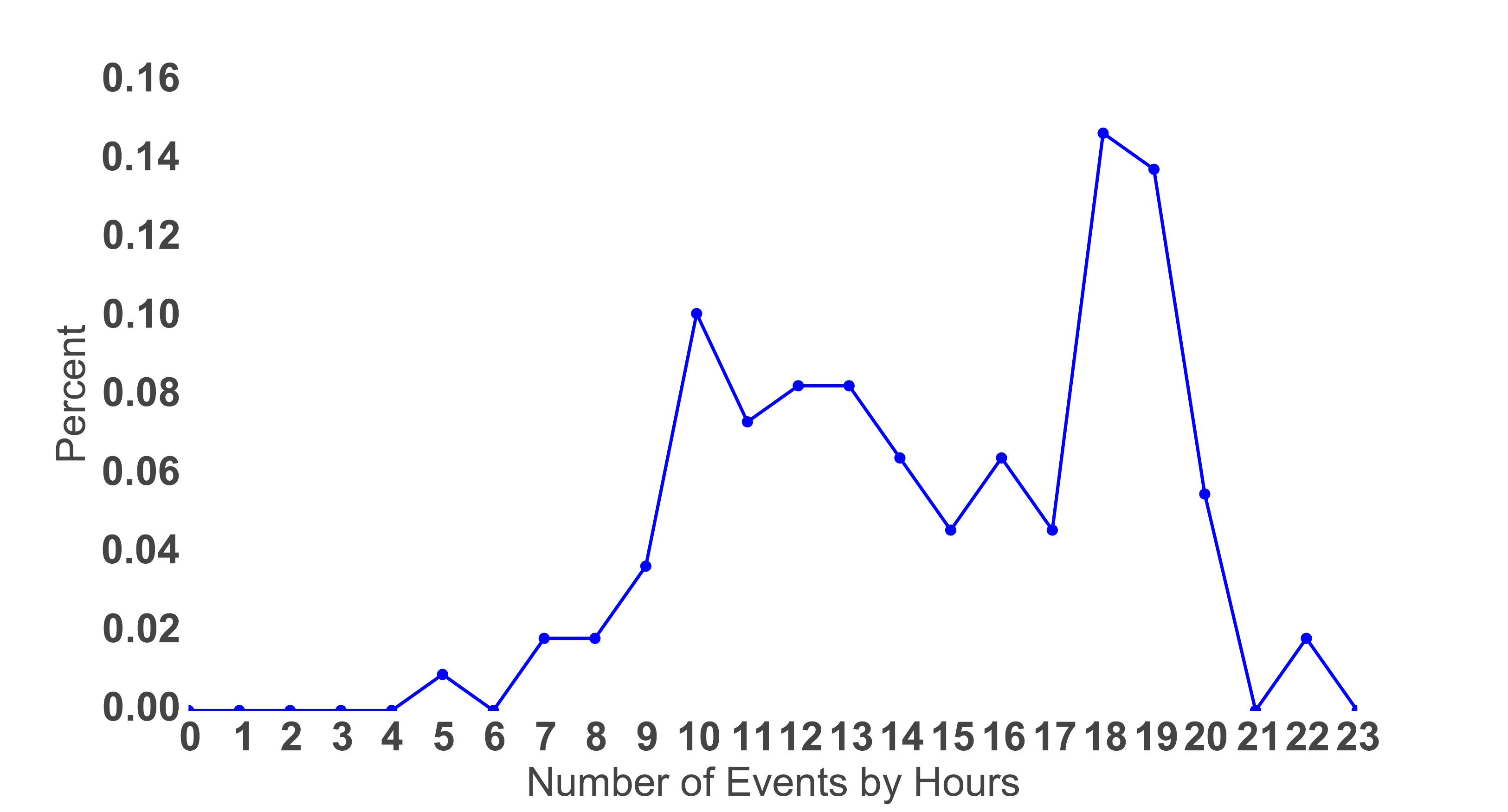}
\caption{The distribution of events by hours on weekday (left) and weekend (right).}
\label{fig:weekday}
\end{figure}

\begin{observation}
People like to attend social events after work on weekdays, while on weekends,
events are distributed relatively evenly.
\end{observation}
Additionally, we are also interested in investigating individual user's temporal preference. Our hypothesis is 
that participants are more likely to attend events after work. Figure~\ref{fig:weekday} depicts the suggested
event times on weekdays and weekends. It is clear that in weekdays there is a high spike around 6pm. While 
in comparison, event times are distributed more evenly throughout the day on weekends. 

\subsection{Impact of Host Preference}
\label{sec:impact-host-preference}
In our OutWithFriendz system, the host has more authority than other
participants. The host can not only decide who to invite, but also finalizes the
event time and location. This suggests that the host will have more influence
on the group decision-making process. In our dataset, we have several
significant observations about host behavior.
\begin{observation}
The final meeting location is closer to a host's frequented place than
other participants.
\end{observation}
It's not surprising that event host would show some ``selfishness'' when making
the final decision. We calculated that the average distance between the final
location and host's closest frequented place is 5.23 miles. While the same
metric for common participants is 6.75 miles, 29\% longer than host, a
significant difference according to a Wilcoxon test ($z=-3.38, p<0.001$).

\begin{table}[]
\centering
\caption{The probability of final event option voted by host and participant}
\label{tab:prob}
\begin{tabular}{|l|c|}
\hline
                                          & Probability \\ \hline
Final event date voted by host            & 0.71        \\ \hline
Final event date voted by participant     & 0.36        \\ \hline
Final event location voted by host        & 0.72        \\ \hline
Final event location voted by participant & 0.34        \\ \hline
\end{tabular}
\end{table}

\begin{observation}
The probability that the final event date and location is voted by the host
is \added{significantly} higher than that for other group participants.
\added{For events in which the host did not choose his/her voting option as the 
final decision, the main reason is to respect the majority voting results.} 
\end{observation}

Table~\ref{tab:prob} shows the probability that final event option is voted for
by the host and by another participant. It is clear that the final option is
much more likely to have been voted for by the host than by other group members,
with a probability of 0.71 vs 0.36 for the final event date ($z=-13.22,
p<0.001$). and 0.72 vs 0.34 ($z=-11.87, p<0.001$) for the final event location.
\added{We also observe that among all the invitations in which the final event time was not 
the host's voting option, 95.2\% coincided with the majority voting results. 
The percentage is 94.4\% for the final event location. This indicates that, although hosts 
have a higher impact on making decisions, they still highly respect
other group members' opinions.}

\begin{observation}
The host choosing not to use the consensus voting result as the final decision
would have negative influence on the event attendance rate.
\end{observation}

In our OutWithFriendz application, the host can select a final decision that is
contrary to the voting results. According to our user study, there are
two main reasons for this behavior: (1) The option that received most votes is
not suitable for the event host; (2) the users discussed through using the app's
chat function and some members changed their minds but did not update their
votes. In our OutWithFriendz dataset, 7.3\% of final dates and 9.2\% of final
locations are contrary to voting results.
We calculated the Pearson correlation between whether the host complies with the
consensus opinion and the corresponding event attendance rate. The results are
shown in Table~\ref{tab:comply}. The positive correlation is significant here for both
location voting and date voting. These results confirm that for event
organization, hosts that don't comply with voting results have negative
impact on the attraction of participants.

\begin{table}[]
\centering
\caption{The correlation between whether host comply voting results and event attendance rate}
\label{tab:comply}
\begin{tabular}{|l|c|c|}
\hline
                                                                      & Pearson Correlation & p-value  \\ \hline
Whether host comply location voting results and event attendance rate & 0.48                & $<10^{-10}$ \\ \hline
Whether host comply date voting results and event attendance rate     & 0.47                & $<10^{-10}$ \\ \hline
\end{tabular}
\end{table}

\subsection{Voting Process Analysis}
\label{sec:voting-process}
Voting is one of the most innovative aspects of our OutWithFriendz system.
In contrast to traditional online event organization services, such as
Meetup and Douban Events, where the meeting location and time is decided only by
group host when the invitation is created, OutWithFriendz allows all group
members to express their preferences through suggestions and votes.
After all invitees have responded to the poll, the group host is able to find a
mutually agreeable location and time that usually accommodates most of the group
members.  Tracking the group's voting process using our system offers a great
opportunity to study group decision making behavior.

\begin{figure}
\centering
\includegraphics[width=0.48\linewidth]{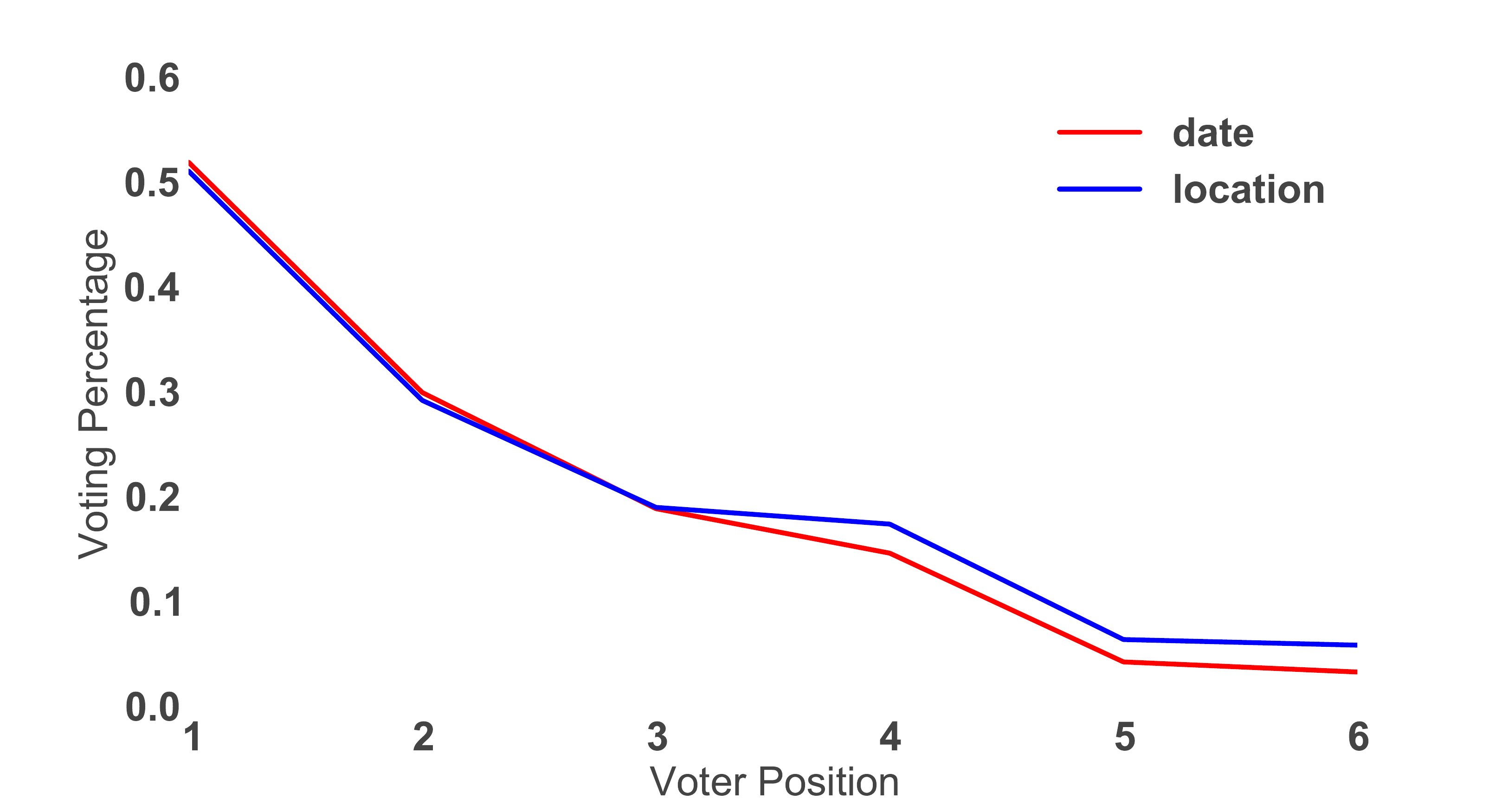}
\caption{The relationship between average availability and voter position.}
\label{fig:votepercentage}
\end{figure}

\begin{observation}
Early voters tend to vote for a wide variety of options, while later voters are
more likely to report limited availability.
\end{observation}

Figure~\ref{fig:votepercentage} shows the relationship between group members'
average availability and their ``voting position''. A user's availability is
defined by Equations~\ref{time_avail} and~\ref{loc_avail}. Voter position refers
to the temporal index of casting votes within the scope of an invitation.
The host's position is 1, the first voter's position is 2, and so on.
There is a clear decrease of availability as voter position increases, and
this result is consistent for both location voting and time voting. There are
several possible explanations for this observation:

\begin{itemize}
\item People who came to the poll later may be busier than early voters, \added{and had 
a smaller time window before the actual event time,}
thus their availability is more limited compared with early voters.
\item The polls in the OutWithFriendz application are all open polls, which
means later voters can see the current voting results. Their votes may
not be able to change the current status significantly because every voter can
only vote once for a given option.
\item Late voters will vote only for agreeable options that help the host to
more easily finalize decisions. This phenomenon will be further discussed by the
next observation.
\end{itemize}

\begin{figure}
\centering
\includegraphics[width=0.48\linewidth]{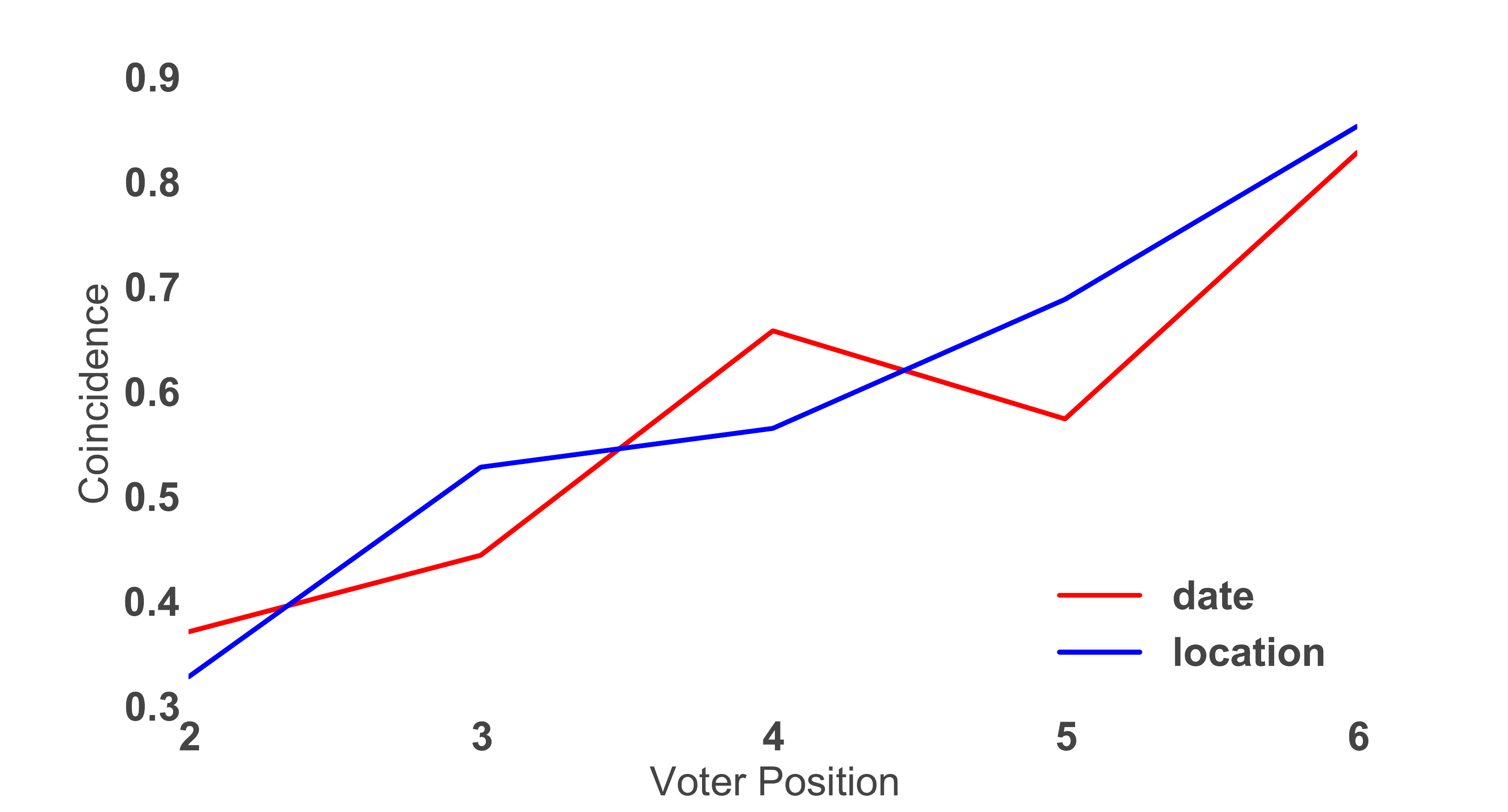}
\caption{The relationship between average voting coincidence and voter position.}
\label{fig:votecoin}
\end{figure}

\begin{observation}
Late voters tend to vote for options that align with existing voting results
and are mutually agreeable.
\end{observation}
Due to the fact that new voters can observe other voters' responses, these early
responses would easily affect future voting behaviors. In our dataset, we find
that later voters are more willing to vote for options that coincide with
existing voting results, which makes it easier for the host to find
common mutually agreeable options.  Here we define the voting coincidence by
cosine similarity:
\begin{equation}
Coincidence = Cosine(\vec{v}, \vec{e})
\end{equation}
where $\vec{v}$ refers to the new voter's voting vector, and $\vec{e}$ refers to
the existing voting vector. For example, if there are four date options in
invitation $i$, and they receive 1, 3, 1, and 0 votes respectively, then the
$\vec{e}$ is $[1, 3, 1, 0]$. If a new voter $v$ votes for second, and fourth
option, then $\vec{v}$ is [0, 1, 0, 1]. The coincidence here is the
cosine similarity between $\vec{v}$ and $\vec{e}$, which is 0.640.
Figure~\ref{fig:votecoin} shows the relationship between average voting
coincidence and voter position. It is clear that there is a positive relationship
between voter position and coincidence in both date voting and location voting.
Later voters will try to consider their options in light of the whole group's
voting behavior.  Sometimes, these later voters may vote for less convenient
options in order to make the host's life easier.

\section{Discussion}

\added{In this section, we summarize the key results of our analysis 
 and provide insights into how these results can be utilized for better 
group event planning experiences. The results cover the impact of user mobility,
individual preference, and the host preference on the group event scheduling process.
We also discuss the impact of user behavior during the voting process.
\newline \textbf{User Mobility.} The analysis in Section
\ref{sec:impact-user-mobility} showed that users with higher mobility are more
likely to be active participants in group events. They are 
more active in voting for proposed event location and event time. We attribute this
behavior to the capability of these users to travel greater distances (perhaps
due to owning a car or having more available time), which gave them flexibility
in selecting location and time options for events. One way to use this
observation for better event planning would be to recommend more diverse
locations and dates for highly mobile users, as they tend to be more willing to
explore new options. On the other hand, users with less mobility should not be
overwhelmed with a large number of choices. In addition to recommending
event locations and time, this observation can be used for forming groups by matching users
with similar mobility levels in the same group, which can lead to smoother event
planning experiences.
\newline \textbf{Individual Preference.} The analysis in Section
\ref{sec:impact-individual-preference} revealed typical patterns related to
individual preferences for event time and locations. First, with regard to event 
location, users tend to arrange events at nearby locations to avoid
traveling long distances. Second, with regard to event time, on weekdays, users want to schedule
events after their working hours, while on weekends users show more
flexibility. These observations are worth considering for smarter group event
planning. For event locations, the application could suggest places such that
the mean travel distance for the group members is minimized, so as to provide a
reasonable compromise for the whole group. Likewise, suggested event time 
should occur outside of typical working hours of the group members.  
\newline \textbf{Host Preference.}  The results in
Section~\ref{sec:impact-host-preference} show that group event planning is
heavily influenced by the host who creates the invitation.  From the analysis in this section,
it is evident that the final event location is on average closer to the host's
frequented locations than that of other group members.  We also see that
the final event locations and dates are more likely to be the options voted by the
host.  However, the influence of the host can also lead to negative outcomes:
when the host chooses not to follow the group's consensus, event attendance is
reduced.  These effects point to the need to carefully consider the preferences
of the host, and how these preferences align with the preferences of the group, when
providing recommendations to event participants.  Effective communication
mechanisms between the host and the participants should also be provided.
\newline \textbf{Voting Analysis.}  The analysis in
Section~\ref{sec:voting-process} shows that the votes cast by early voters are
very likely to affect late voters.  Late voters tend to vote for fewer options,
and these options tend to match those that have already received votes from
early voters.  This phenomenon can be used to improve the event planning
experience.  For example, we could encourage users to vote early, so that their
votes will carry more weight.  We could also hide existing voting results, so as
to prevent existing votes from biasing later voters, and facilitate the voting
process by providing voting recommendations to users based on their historical
voting patterns.
}

\deleted{\textbf{Expanding event coverage.} Though we have collected more
than 300 completed group events, we hope to grow our user base through more
effective advertising, so that we may achieve viral adoption and gather data at
even larger scales.}  



\deleted{\textbf{Facebook friends requirement}.  To obtain a user's friend list, OutWithFriendz currently only allows users to log in through their Facebook accounts. Users may also want to invite people who are not already a Facebook friend or do not use Facebook at all. This limits our app's ability to support larger groups. We plan to design an ``Add Friend' function which enables users to log in and connect with other users directly within the application.}

\deleted{\textbf{Different voting mechanisms.} All the polls in OutWithFriends are
currently designed to be open polls, which means late-coming voters can see
existing voting results, which may influence their vote.  In a future version of
the application, we plan to add a closed poll option. For closed polls, existing
voting results will be hidden from new voters. This functionality will allow us
to examine how a closed poll mechanism influences the group decision-making
process.}

\deleted{\textbf{Group recommendation.} 
We have seen from our work on OutWithFriendz that a number of factors influence
group decision making behavior. We believe that group context could be seen as
inhabiting a latent trait space, similar to how users inhabit a latent user
trait space in the matrix factorization framework for individual recommendation.
Furthermore, our work has revealed that host and individual influence within a
group plays a significant role in the group decision-making process. In future
work, we intend to pursue the development of a group recommendation system that
incorporates these ideas into a probabilistic model for group preference
behavior. Building a distributed system implementation of mobile group
recommendation will help us gain a better understanding of mobile group
dynamics in the real world, and provide useful suggestions for group organizers.}

\section{Conclusion and Future Work} 
\label{sec:conclusion} 

In this paper, we have presented the results of a large-scale study of OutWithFriendz,
a newly designed mobile application for group event \replaced{scheduling }{organization}. We summarize
our key findings as follows: (1) User mobility has a significant impact on group event attendance.
(2) Users would like to vote for locations near their frequented places. On weekdays, they would 
like to meet after work while on weekends, they have wider time options. (3) A group host 
has a higher impact on the group decision making process than other members. (4) Early
voters are more likely to vote for more options while late voters tend to coincide with existing
results to find a mutually agreeable option. We believe the results presented in this paper
are a good start towards \replaced{better }{accurate} understanding of group \added{event scheduling} behaviors in real life. Our analysis
of this first-of-a-kind real world user study of mobile group dynamics has provided very 
important guidance for \replaced{group event scheduling and recommendation }{mobile group recommendation and mobile group formation.}

\added{We plan to pursue several directions for future work. First,
although we have collected more than 300 completed group events, we hope to grow
our user base through more effective advertising, so that we may achieve viral
adoption and gather data at even larger scales. Second, to obtain a user's
friend list, OutWithFriendz currently only allows users to log in through their
Facebook accounts. Users may want to invite people who are not already a
Facebook friend or do not use Facebook at all. This limits our app's ability to
support larger groups. We plan to design an ``Add Friend'' function which enables
users to log in and connect with other users directly within the application.
Third, currently, polls in OutWithFriends are designed to be open
polls, allowing late-coming voters to see existing voting results, which may
influence their own votes. We plan to add a closed poll option. For closed polls,
existing voting results will be hidden from new voters. This functionality will
allow us to examine how a closed poll mechanism influences the group
event scheduling process. Lastly, we have seen from our work on OutWithFriendz
that a number of factors influence group decision making. We believe that
group context can be seen as inhabiting a latent trait space, similar to how
users inhabit a latent user trait space in the matrix factorization framework
for individual recommendation. Furthermore, our work has revealed that both host and
individual members within a group play an important role in the group
event scheduling process. In our future work, we intend to pursue the development of
a group recommendation system that incorporates these ideas into a probabilistic
model for group preferences and make group event recommendations in real-world 
settings. 
This will help us gain a better understanding of group event dynamics and provide useful
suggestions for group event organizers.}

\bibliographystyle{ACM-Reference-Format}
\bibliography{paper}

\end{document}